\documentclass[11pt,twoside]{article}
\usepackage{epsfig}
\renewcommand\refname{References}

\renewenvironment{thebibliography}[1]
 { \section*{\refname}
		\begin{list}{\arabic{enumi}.}
    {\usecounter{enumi} \setlength{\parsep}{0pt}
     \setlength{\itemsep}{3pt} \settowidth{\labelwidth}{#1.}
     \sloppy
    }}{\end{list}}

\begin{document}

\title{Quantum mechanics on Hilbert manifolds: 
The principle of functional relativity }
\author{Alexey A. Kryukov
\footnote{Department of Mathematics, University of Wisconsin Colleges
\newline
E-mail: alexey.kryukov@uwc.edu, aakrioukov@facstaff.wisc.edu}
}
\maketitle

\pagestyle{myheadings} 
\thispagestyle{plain}         
\markboth{Alexey A. Kryukov}{QM on Hilbert manifolds} 
\setcounter{page}{1}  

\vskip30pt
\begin{abstract}
Quantum mechanics is formulated as a geometric theory on a Hilbert manifold.
Images of charts on the manifold are allowed to belong to arbitrary Hilbert spaces of functions including spaces of generalized functions. Tensor equations in this setting, also called {\em functional tensor equations}, describe families of functional equations on various Hilbert spaces of functions. 
The {\em principle of functional relativity} is introduced which states that quantum theory
is indeed a functional tensor theory, i.e., it can be described by 
functional tensor equations. The main equations of quantum theory are shown to be compatible with the principle of functional relativity. By accepting the principle as a hypothesis, we then explain the origin of physical dimensions, provide a geometric interpretation of Planck's constant, and find a simple interpretation of the two-slit experiment and the process of measurement. 
\end{abstract}

\section{ Introduction}
\setcounter{equation}{0}

One of the most important goals of modern theoretical physics is to reconcile two of its cornerstones: general relativity (GR) and quantum theory (QT). Both theories have been extremely powerful and precise in explaining and predicting the observed phenomena. Accordingly, both theories are expected to be present in some way in any future theory. The areas of applicability of general relativity (also called the theory of gravitation) and quantum theory  are, in a way, opposite. The quantum theory is an ultimate theory of the world of microscopic particles and fields, while general relativity deals primarily with objects and processes of a macroscopic character. 

The theories seem to be dissimilar and incompatible in every possible way. This becomes clear already when comparing the mathematical machinery used in each theory. Roughly speaking, the quantum theory is described in terms of linear operators in Hilbert spaces with a heavy use of functional methods and representation theory. At the same time, general relativity is based on the finite dimensional Riemannian geometry and uses primarily the methods of differential geometry and partial differential equations. In simple words, the world of quantum theory is infinite-dimensional and primarily linear, while the world of general relativity is finite dimensional and non-linear.

The theory of gravitation is naturally {\em local}, that is, physical observations at a point in the theory depend only on the state of matter and fields in the immediate neighborhood of the point. Mathematically this is reflected in the fact that the equations of gravitation are partial differential equations. The quantum theory is also local as it is also described by means of differential equations. However, the locality of quantum theory does not work that well and seems to be imposed upon us by the lack of a better mathematical description. In particular, many of the difficulties in the quantum field theory (QFT) seem to be rooted in the concept of a field at a point in space-time. This concept seems to be both necessary and contradictory leading to divergences in the theory.

Some of the difficulties of QFT are also present in non-relativistic quantum mechanics (QM) in the form of the so-called {\em improper states}.
The latter are the states, like the eigenstates of position and momentum operators, that are non-square integrable and, as a result, do not always fit nicely into the theory. At the same time, the improper states are essential as they serve as the building blocks of the quantum theory and simultaneously provide the link between the quantum and the classical worlds. Indeed, the state function in QM would not be defined without our ability to measure positions of non-relativistic particles. Likewise, the scattering amplitude in QFT would not exist without our ability to measure momenta of free particles. Simultaneously, the latter measurements ideally create improper states thereby endowing the particles with the classical mechanical properties and providing the foundation of the classical world. 

The mathematical difficulties related to the presence of improper states in QM are usually resolved by approximating these states, in some way, by square-integrable functions. Alternatively, the improper states can be rigorously defined as functionals in the rigged Hilbert space construction of Gel'fand (see Ref. 1), in which case they have no norm. Both approaches make the theory somewhat awkward as the improper states, being  the building blocks of the theory, are not then included in the theory on an equal footing with the square-integrable states. Moreover, the latter mathematical fact is but one indication that the quantum theory, while based on the classical properties of matter, is unable to fully explain these properties. Numerous other observations, both theoretical and experimental, all seem to be leading to the same conclusion of incompleteness of quantum theory. This incompleteness persists also in the advanced forms of quantum theory such as the string/M theory, which rely on a pre-existing notion of classical space-time. Formulating the quantum theory in a way independent of the pre-existing classical space and of the classical properties of measuring devices becomes then a problem of fundamental importance.
In light of the properties of general relativity and quantum theory discussed above, the problem expressed in a very general way consists in deriving the ``finite dimensional nonlinear world'' from the ``infinite-dimensional linear one''. 

In a recent work (Refs. 6, 7) improper states in quantum mechanics have been put on an equal footing with square-integrable states by means of a {\em functional coordinate formalism on Hilbert manifolds}. 
The coordinate charts on a Hilbert manifold in the formalism take values in arbitrary infinite-dimensional separable Hilbert spaces of functions including spaces of generalized functions. Isomorphisms of these spaces are then identified with transformations of coordinates on the manifold. The resulting formalism generalizes the notion of a tensor and seems to be the most appropriate and powerful extension of the local coordinate approach to tensor fields to the case of infinitely many dimensions. The formalism demonstrates, in particular, that the improper states can be naturally included in QT if one is ready to accept that the Hilbert metric on the space of states can have a different functional form in different coordinate charts and in different physical situations. 

Furthermore, in Ref. 8 the local coordinate formalism of finite dimensional Riemannian geometry has been naturally derived from the above functional coordinate formalism on Hilbert manifolds. This opened a way of reformulating the Riemannian geometry, topology and physics of classical space-time in functional terms. In fact, the geometry of the classical space itself as well as the dynamics of classical and quantum particles on the space have been {\em derived} in Ref. 8 from the geometry of a Hilbert space of functions of abstract parameters. To put it differently, the geometry of the classical space and the dynamics of particles on the space have been shown to be ``encoded'' into the geometry of an appropriate Hilbert space of functions of abstract parameters. In particular, the formalism eliminates the need for a pre-existing classical space in quantum theory.

The apparent success of the above formalism in bridging the gap between the quantum and the classical worlds supports the idea that Hilbert manifolds offer an appropriate arena while the formalism itself provides an appropriate mathematical language for quantum physics. At the same time, the resulting extension of the currently accepted space-time arena is, in a way, minimal. In fact, the quantum theory already uses various infinite-dimensional Hilbert spaces as an essential part of its formalism. The obtained results simply hint that Hilbert spaces and, more generally, Hilbert manifolds  should play an even larger role in modern physics.

In the current paper we continue developing the above mentioned geometric approach by exploring the idea that quantum theory is a functional tensor theory. In other words, the equations of quantum theory can be expressed in a form independent of any particular functional realization. 
This constitutes what is called in the paper the {\em principle of  functional relativity}. We show that the principle is a natural extension of the classical principle of relativity on space-time. Simultaneously, the principle is in apparent agreement with the standard apparatus of quantum theory.  
By accepting the principle as a hypothesis, we explain the origin of physical dimensions, provide a geometric interpretation of Planck's constant, and find a simple model of the two-slit experiment and the process of measurement.

Here is a plan of the paper. In Sec. 2 we briefly review the previously obtained results  
concerning the functional coordinate formalism and its applications in quantum theory. 
In Sec. 3 we relate the observables in QM with vector fields in a Hilbert space and prepare the ground for a geometric interpretation of QM. 
In Sec. 4 we introduce a Riemannian metric on the unit sphere $S^{L_{2}}$ in a Hilbert space $L_{2}$ of square-integrable state functions and in the corresponding projective space $CP^{L_{2}}$ and verify that the integral curves of the vector fields associated with observables are geodesics in this metric.
A simpler but similar analysis is done in Sec. 5 where we discuss the Killing metric on the sphere $S^{3}$ of unit spinors and the Fubini-Study metric on the complex projective space $CP^{1}$ of physical spinors.  
The principle of functional relativity is introduced in Sec. 6. Here we show that the apparatus of quantum theory is consistent with the principle of functional relativity, that classical relativity is a special case of functional relativity and that the speed of light is a functional scalar. 
In Sec. 7 we use the principle of functional relativity to investigate the origin of physical dimensions and of quantum commutators. In particular, the commutators in quantum theory are related to the curvature of the Riemannian manifold $S^{L_{2}}$. 
The process of measurement in QM is analyzed in Sec. 8. Here possible interpretations of the two-slit experiment and of the instantaneous nature of collapse in light of the principle of functional relativity are proposed and future applications of the theory are discussed.

\section{Functional coordinate formalism on Hilbert manifolds}

\setcounter{equation}{0}

The paper will make an extensive use of the coordinate formalism on Hilbert manifolds developed in Refs. 6-8. The readers is referred to Ref. 9 for a mathematically rigorous introduction to the formalism and its applications. The main idea of the formalism is to associate a specific functional form of physical quantities (e.g., observables, states, etc.) in QT with realization in a particular Hilbert space of the corresponding invariant quantities defined on an abstract Hilbert space.   

For instance, the (pure) state of a quantum system in standard QM is defined in terms of state function, which is an element of a particular Hilbert space. This is similar to defining a point in space-time as a $4$-tuple of coordinates. 
The $4$-tuple may pick out a space-time point, but it cannot be identified with the point because there are other ways of picking it out. The point itself is a {\em geometric} object, which is independent of any particular coordinates. A quantum state can be defined in a similar geometric way. In the paper the state is considered as a point in an abstract state space, called a {\em string space} and the state function in a particular Hilbert space is interpreted as a kind of ``coordinate-dependent'' way of picking out a state. We remark that, except for the shared general infinite-dimensional setting, the ``string'' formalism developed here has nothing to do with either string theory or loop gravity. 
Here are the main definitions:

A {\em string space} $\bf {S}$ is an abstract
infinite-dimensional linear topological space isomorphic (that is, topologically linearly isomorphic) to a separable Hilbert space.
The elements of $\bf {S}$ are called strings and will be denoted by the capital Greek letters
$\Phi,\Psi,...\quad.$

A {\em Hilbert space of functions} (or a {\em coordinate space}) is either a Hilbert space
$H$, elements of which are equivalence classes of maps between two given
subsets of $R^{n}$ or the Hilbert space $H^{\ast }$ dual to $H$. 
In other words, each equivalence class of either $H$ or $H^{\ast}$ contains a representative which is a numeric or a vector-valued function of $n$ variables or a functional on such functions. We remark here that the number of variables $n$ may vary from space to space. 

A linear isomorphism ${e}_{H}$ from a Hilbert space $H$
of functions onto $\bf {S}$
is called a {\em string basis} (or a {\em functional basis}) on $\bf {S}$. The inverse map ${e}^{-1}_{H}: {\bf S} \longrightarrow H$ is called a {\em linear coordinate system on ${\bf S}$} (or a {\em linear functional coordinate system}).
The string basis identifies a string with a function: if $\Phi \in {\bf S}$, then ${\Phi }={e}_{H}(\varphi)$ for a unique $\varphi \in H$.

Let ${\bf {S}}^{\ast}$ be the dual string space. That is, ${\bf {S}}^{\ast}$
is the space of all linear continuous functionals on strings. Likewise, let $H^{\ast}$ be the dual of a coordinate space $H$.
A linear isomorphism ${e}_{H^{\ast }}$ of $H^{\ast }$ onto ${\bf {S}}^{\ast}$
is called a {\em string basis on} ${\bf {S}}^{\ast}$.

The basis ${e}_{H^{\ast }}$ is called {\em dual} to the
basis ${e}_{H}$ if for any string ${\Phi }={e}_{H}(\varphi)$ and for any
functional ${F}={e}_{H^{\ast }}(f)$ in ${\bf {S}}^{\ast}$ the following is true:
\begin{equation}
{F}({\Phi })=f(\varphi ).
\end{equation}

In the future the action of a linear functional $f$ on function $\varphi$ will be denoted in one of the following three ways: $f(\varphi)=(f,\varphi)=(\varphi, f)$.
The expressions like $(f,\varphi)$ will be distinguished from the inner product of two elements in a Hilbert space $H$ by the subscript $H$ in the symbol of inner product. For instance, if $f$, $g$ are elements of $H$, then their inner product will be denoted by $(f,g)_{H}$.


By definition the string space $\bf {S}$ is isomorphic to a
separable Hilbert space. We can furthermore assume that $\bf {S}$ itself is an abstract Hilbert space. 
Accordingly, we will assume that the string bases
$e_{H}$ are isomorphisms of Hilbert spaces. That is,
the Hilbert metric on any coordinate space $H$ is determined by the Hilbert metric on $\bf {S}$ and the choice of a string basis. Conversely, the choice of a coordinate Hilbert space determines the
corresponding string basis $e_{H}$ up to a unitary transformation. Indeed, with $H$ fixed, any two bases $e_{H}$, 
${\widetilde e}_{H}$ can only differ by an automorphism of $H$, i.e., by a unitary transformation.

Assume for simplicity that $H$ is a real Hilbert space (generalization to the case of a complex Hilbert space will be obvious). We have:
\begin{equation}
\label{orto}
(\Phi ,\Psi )_{S}={\bf{G}}(\Phi ,\Psi )=G(\varphi,\psi
)=g_{kl}\varphi ^{k}\psi ^{l},
\end{equation}
where ${\bf {G}}:{\bf {S}}\times {\bf {S}}\longrightarrow R$
is a bilinear form defining the inner
product on $\bf {S}$ and $G:H\times H\longrightarrow R$ is the induced bilinear form. The expression on the right is a convenient
form of writing the action of $G$ on $H\times H$. Such an {\em index notation} will be useful in the paper.

A string basis ${e}_{H}$ in $\bf {S}$ will be called
{\em orthogonal} if for any $\Phi, \Psi \in {\bf S}$ we have
\begin{equation}
(\Phi ,\Psi )_{S}=f_{\varphi}(\psi ),
\end{equation}
where $f_{\varphi}$ is a {\it regular} functional and $\Phi=e_{H}\varphi$,
$\Psi=e_{H}\psi$ as before. That is,
\begin{equation}
\label{orto1}
(\Phi,\Psi)_{S}=f_{\varphi}(\psi)=\int \varphi(x)\psi(x)d\mu (x),
\end{equation}
where $\int$ here denotes an actual integral over a $\mu $-measurable set $D \in R^{n}$ which is the domain of definition of functions in $H$.

If the integral in Eq. (\ref{orto1}) is the usual Lebesgue integral and/or a sum over a discrete index $x$, the corresponding coordinate space will be called an $L_{2}$-space. In this case we will also say that the basis $e_{H}$ is {\em orthonormal}.
If the integral is a more general Lebesgue-Stieltjes integral, the coordinate space defined by Eq. (\ref{orto1}) will be called an $L_{2}$-space with the weight $\mu$ and the basis $e_{H}$ will be called {\em orthogonal}.
Roughly speaking, the metric on Hilbert spaces defined by orthogonal string bases has a ``diagonal'' kernel. In particular, the kernel may be proportional to the delta-function or to the Kr{\oe}necker symbol. More general coordinate Hilbert spaces have a ``non-diagonal'' metric (see Eq. (\ref{example_H}) for example).

The bilinear form ${\bf{{G}:{S}}}\times {\bf {S}}\longrightarrow R$
generates a linear isomorphism $\widehat{\bf{G}}:{\bf {S}}
\longrightarrow {\bf{{S}}^{\ast }}$ by ${\bf{G}}(\Phi ,\Psi )=(\widehat{
\bf{G}}\Phi ,\Psi )$.
In any basis $e_{H}$ we have
\begin{equation}
(\Phi ,\Psi )_{S}=(\widehat{\bf{G}}e_{H}\varphi ,e_{H}\psi )=e_{H}^{\ast
}\widehat{\bf{G}}e_{H}\varphi (\psi )={\widehat G} \varphi (\psi),
\end{equation}
where $e_{H}^{\ast}$ is the adjoint of $e_{H}$ and ${\widehat G}=e_{H}^{\ast }\widehat{\bf{G}}e_{H}$ maps $H$ onto $H^{\ast }$. Here the adjoint of a linear operator ${\widehat A}:H \longrightarrow {\widetilde H}$ is the operator ${\widehat A}^{\ast}:{\widetilde H}^{\ast}\longrightarrow H^{\ast}$  defined by $({\widehat A}^{\ast}f, \varphi)=(f, {\widehat A}\varphi)$ for any $\varphi$ in $H$ and any $f$ in ${\widetilde H}^{\ast}$.
If $e_{H}$ is orthogonal, then ${\widehat G}\varphi
=f_{\varphi} $.
It follows from the definition that if $e_{H}$ is orthogonal, then $H$ is a space $L_{2}(D,\mu)$ of square-integrable functions on a $\mu$-measurable set $D \in R^{n}$.
In particular, not every coordinate Hilbert space $H$ can
produce an orthogonal string basis ${e}_{H}$. 

Let us remark that the above definitions are analogous to their finite dimensional counterparts. In fact, in the case of a finite number of dimensions the definition of a string space becomes simply the definition of an abstract $n$-dimensional vector space $V$. A string basis becomes a map from the space $R^{n}$ of $n$-tuples onto $V$ and can be identified with the ordinary basis on $V$. Likewise, the dual string basis becomes a basis dual to the ordinary basis. A similar ``correspondence rule'' is valid for all of the above definitions.
At the same time, in the infinite-dimensional case the given definitions describe substantially new objects. The main property of these objects is their invariance under various isomorphisms of Hilbert spaces of functions. 

In particular, it is important to distinguish clearly the notion of a string basis from the notion of an ordinary basis on a Hilbert space.
Namely, a  string basis permits us to represent invariant objects in string space (strings) in terms of functions, which are elements of a Hilbert space of functions. A basis on the space of functions then allows us
to represent functions in terms of numbers; that is, in terms of the components
of the functions in the basis. As already discussed, in case of a finitely many dimensions the difference disappears.

By a {\em linear coordinate transformation on $\bf {S}$} we understand 
an isomorphism $\omega :\widetilde{H}\longrightarrow H$ of Hilbert
spaces which defines a new string
basis $e_{\widetilde{H}}:\widetilde{H}\longrightarrow \bf {S}$ by
$e_{\widetilde{H}}=e_{H}\circ \omega $.

Let $\varphi=e^{-1}_{H}\Phi $, ${\widehat A}=e^{-1}_{H}{\bf {\widehat A}}e_{H}$ and ${\widehat G}=\left( e^{-1}_{H}\right)^{\ast}{\bf {\widehat G}}e^{-1}_{H}$ be the coordinate expressions of a string $\Phi$, an operator ${\bf {\widehat A}}: {\bf S} \longrightarrow {\bf S}$ and the metric $\widehat{\bf{G}}:{\bf {S}}
\longrightarrow {\bf{{S}}^{\ast }}$ in a basis $e_{H}$. Let $\omega :\widetilde{H}\longrightarrow H$ be a linear coordinate transformation on ${\bf S}$.
Then we easily obtain the following transformation laws:
\begin{eqnarray}
\label{tensor}
\varphi &=&\omega\widetilde{\varphi}\\
\label{tensor1}
\widehat{G}_{\widetilde{H}} &=&\omega^{*}\widehat{G}\omega\\
\label{tensor3}
{\widehat A}_{\widetilde{H}} &=&\omega^{-1}{\widehat A}\omega,
\end{eqnarray}
where $\widetilde{\varphi}$, ${\widehat A}_{\widetilde{H}}$ and $\widehat{G}_{\widetilde{H}}$ are coordinate functions of $\Phi$, ${\bf {\widehat A}}$ and $\widehat{\bf{G}}$ in the basis $e_{\widetilde{H}}$.

More generally, consider an arbitrary Hilbert manifold ${\it S}$ modeled on ${\bf S}$.
Let $(U_{\alpha },\pi _{\alpha })$ be an atlas on $\it {S}$ (i.e. a collection of opens sets $U_{\alpha }$ covering ${\it S}$ and diffeomorphisms $\pi _{\alpha }$ of $U_{\alpha }$ onto subsets of ${\bf S}$).
A collection of quadruples $(U_{\alpha },\pi _{\alpha
},\omega _{\alpha },H_{\alpha })$, where each $H_{\alpha }$ is a Hilbert
space of functions and $\omega _{\alpha }$ is an isomorphism of $\bf {S}$
onto $H_{\alpha }$ is called a {\em functional atlas} on
$\it {S}$. A collection of all compatible functional atlases on $\bf {S}
$ is called a {\em coordinate structure} on $\it {S}$. A Hilbert manifold ${\it S}$ with the above coordinate structure is called a {\em string manifold} or a {\em functional manifold}.

Let $(U_{\alpha },\pi _{\alpha })$ be a chart on $\it {S}$.
If $p\in U_{\alpha },$ then
$\omega _{\alpha }\circ \pi _{\alpha }(p)$ is called the {\em coordinate}
of $p$. The map $\omega _{\alpha }\circ \pi _{\alpha }:U_{\alpha} \longrightarrow H_{\alpha}$ is called a {\em coordinate system}. The isomorphisms
$\omega _{\beta }\circ \pi _{\beta }\circ (\omega _{\alpha }\circ
\pi _{\alpha })^{-1}:\omega _{\alpha }\circ \pi _{\alpha }
(U_{\alpha }\cap U_{\beta })\longrightarrow
\omega _{\beta }\circ \pi _{\beta }(U_{\alpha }\cap U_{\beta })$ are called
{\em string (or functional) coordinate transformations}.

As $\it {S}$ is a differentiable manifold one can also introduce the tangent
bundle structure $\tau :T{\it{S}}\longrightarrow {\it {S}}$ and the bundle $\tau
_{s}^{r}:T_{s}^{r}{\it {S}}\longrightarrow {\it {S}}$
of tensors of rank $(r,s)$. Whenever necessary to distinguish tensors (tensor fields) on ordinary Hilbert manifolds from tensors on string manifolds, we will call the latter tensors the {\em string tensors} or the {\em functional tensors}. Accordingly, the equations invariant under string coordinate transformations will be called the {\em string tensor} or the {\em functional tensor equations}. 

A coordinate structure on a Hilbert manifold permits one to obtain a functional
description of any string tensor. Namely, let ${\bf{G}}_{p}(F_{1},...,F_{r},\Phi
_{1},...,\Phi _{s})$ be an $(r,s)$-tensor on $\it {S}$.
The coordinate map $\omega _{\alpha }\circ \pi _{\alpha
}:U_{\alpha }\longrightarrow H_{\alpha }$ for each $p\in U_{\alpha }$\ yields
the linear map of tangent spaces $d\rho _{\alpha }:T_{\omega _{\alpha }\circ
\pi _{\alpha }(p)}H_{\alpha }\longrightarrow T_{p}\it {S}$, where $\rho
_{\alpha }=\pi _{\alpha }^{-1}\circ \omega _{\alpha }^{-1}.$ This map is
called a {\it local coordinate string basis on} $\it {S}$.
Notice that for each $p$ the map\ $e_{H_{\alpha }} \equiv e_{H_{\alpha }}(p)$ is a string basis as
defined earlier. Therefore, the local dual basis $e_{H_{\alpha
}^{\ast }}=e_{H_{\alpha }^{\ast }}(p)$ is defined for each $p$
as before and is a function of $p.$

We now have $F_{i}=e_{H_{\alpha }^{\ast }}f_{i}$, and $\Phi
_{j}=e_{H_{\alpha }}\varphi _{j}$ for any $F_{i}\in T_{p}^{\ast }\it {S}$
, $\Phi _{j}\in T_{p}\it {S}$ and some $f_{i}\in H_{\alpha }^{\ast
},\varphi _{j}\in H_{\alpha }$. Therefore the equation
\begin{equation}
{\bf{G}}_{p}(F_{1},...,F_{r},
\Phi _{1},...,\Phi _{s})=G_{p}(f_{1},...,f_{r},\varphi _{1},...,\varphi _{s})
\end{equation}
defines component functions of the $(r,s)$-tensor ${\bf{G}}_{p}$ in the
local coordinate basis $e_{H_{\alpha }}$.

The outlined functional coordinate formalism permits one to consider Hilbert spaces containing singular generalized functions on an equal footing with spaces of square-integrable functions. In fact, consider a Hilbert space $H$ of functions finite in the metric associated with the inner product
\begin{equation}
\label{H_metric}
(\varphi,\psi)_{H}=\int k(x,y)\varphi(x)\psi(y)dxdy.
\end{equation}
In Eq. (\ref{H_metric}) the kernel $k(x,y)$ is an appropriate function on, say, $R^{n} \times R^{n}$ and the integral sign is understood as the action of the corresponding bilinear functional on $H \times H$.
More constructively, $H$ can be obtained by completing a space of ordinary functions $\varphi$ with respect to the norm $\left\|\varphi\right\|_{H}^{2}=(\varphi, \varphi)_{H}$. We remark here that only those functions $k(x,y)$ for which Eq. (\ref{H_metric}) is a non-degenerate inner product (i.e. the corresponding completion $H$ is a Hilbert space) are considered.

By changing the ``smoothness" properties of $k(x,y)$ as well as its behavior at infinity we change the variety of functions in $H$. If, for example, the kernel $k(x,y)$ is a smooth function, then the corresponding Hilbert space contains various singular generalized functions.
In particular, the space $H$ of real valued generalized functions ``of"  (i.e. defined on functions of) $x \in R^{n}$  finite in the metric
\begin{equation}
\label{example_H}
(\varphi,\psi)_{H}=\int e^{-(x-y)^{2}}\varphi(x)\psi(y)dxdy
\end{equation}
can be shown to be Hilbert (see Ref. 6). 
Such a space contains the delta-functions as, for example,
\begin{equation}
\label{HHH}
\int e^{-(x-y)^{2}}\delta(x)\delta(y)dxdy=1.
\end{equation}
Moreover, $H$ contains the derivatives of any order of the delta-functions as well. 

By allowing for generalized functions to be elements of a Hilbert space of states it becomes possible to extend to such functions the standard QM formalism dealing with square-integrable functions. For instance, the expectation value of position observable ${\widehat x}$ for a particle in position eigenstate $\delta_{a}(x)=\delta(x-a)$ in the space $H$ with metric Eq. (\ref{example_H}) is
\begin{equation}
(\delta_{a}, {\widehat x}\delta_{a})_{H}=\int e^{-(x-y)^{2}}\delta(x-a)y\delta(y-a)dxdy=a.
\end{equation}

Although this result makes perfect sense, the expectation value $\left(\varphi, {\widehat x}\varphi \right)_{H}$ for a square integrable function or a superposition of delta-functions will be only approximately equal to what one would expect from the standard QM. The same is true about more general bilinear expressions. A nice resolution of this problem will be given in Sec. 7.  

Let us also illustrate the usefulness of string tensor equations and their difference from the ordinary tensor equations. For this let us consider the {\em generalized eigenvalue problem} 
\begin{equation}
\label{LLLL}
F({\bf{\widehat A}}\Phi)=\lambda F(\Phi),
\end{equation}
for a linear operator ${\bf {\widehat  A}}$ on ${\bf S}$. The problem consists in finding all functionals $F \in {\bf S}^{\ast}$ and the corresponding numbers $\lambda$ for which the string tensor equation Eq. (\ref{LLLL}) is satisfied for all $\Phi \in {\bf S}$.

Assume that the pair $F, \lambda$ is a solution of Eq. (\ref{LLLL}) and $e_{H}$ is a string basis on ${\bf S}$. Then we have
\begin{equation}
\label{1111}
e_{H}^{\ast }F(e_{H}^{-1}{\bf{\widehat A}}e_{H}\varphi )=\lambda e_{H}^{\ast
}F(\varphi ),
\end{equation}
where $e_{H}\varphi =\Phi $ and
$e_{H}^{-1}{\bf{\widehat A}}e_{H}$ is the representation of ${\bf{\widehat A}}$ in
the basis $e_{H}$.
By defining $e_{H}^{\ast }F=f$ and ${\widehat A}=e_{H}^{-1}{\bf{\widehat A}}e_{H}$, we have
\begin{equation}
\label{eigen_new}
f({\widehat A}\varphi )=\lambda f(\varphi ).
\end{equation}
Notice that the last equation describes not just one eigenvalue problem, but
a family of such problems, one for each string basis $e_{H}$. As we change
$e_{H}$, the operator $A$ in general changes as
well, as do the eigenfunctions $f$. 

For instance, 
let $H \subset L_{2}(R)$ be a Hilbert space of complex-valued functions such that the action of the operator of
differentiation ${\widehat A}=-i\frac{d}{dx}$ is defined on $H$ and the dual space $H^{\ast}$ contains the functionals $f(x)=e^{ipx}$. For example, the Hilbert metric on $H^{\ast}$ could be given by the kernel $e^{-\frac{x^{2}}{2}}\delta(x-y)$ (see Sec. 6).
The generalized eigenvalue problem
for ${\widehat A}$ is
\begin{equation}
\label{eigenvalue1}
f\left (-i\frac{d}{dx}\varphi \right )=p f\left (\varphi \right ).
\end{equation}
The equation Eq. (\ref{eigenvalue1}) must be satisfied for every $\varphi$ in $H$. The functionals
\begin{equation}
f(x)=e^{ip x}
\end{equation}
are the eigenvectors of $A$. Let us now consider the coordinate transformation $\rho:
H \longrightarrow \widetilde{H}$ given by the Fourier transform:
\begin{equation}
\psi (k)=(\rho \varphi )(k)=\int \varphi (x)e^{ikx}dx.
\end{equation}
The Fourier transform induces a Hilbert structure on the space ${\widetilde H}=\rho (H)$. Relative to this structure $\rho$ is an isomorphism of the Hilbert spaces $\widetilde{H}$ and $H$. The inverse transform
is given by
\begin{equation}
(\omega \psi )(x)=\frac{1}{2\pi }\int \psi (k)e^{-ikx}dk.
\end{equation}
Notice that the Fourier transform of $e^{ipx}$ is $\delta (k-p)$ and therefore the space dual to $\widetilde{H}$ contains delta-functions. In particular, if the kernel of the metric on $H^{\ast}$ is given by $e^{-\frac{x^{2}}{2}}\delta(x-y)$, then the metric on ${\widetilde H}^{\ast}$ has the kernel proportional to $e^{-\frac{1}{2}(x-y)^{2}}$ (see Sec. 6).
According to Eq. (\ref{1111}), the generalized eigenvalue problem in new coordinates is
\begin{equation}
\omega ^{\ast }f(\rho {\widehat A}\omega \psi )=p \omega ^{\ast }f(\psi ).
\end{equation}
We have:
\begin{equation}
{\widehat A}\omega \psi =-i\frac{d}{dx}\frac{1}{2\pi }\int \psi (k)e^{-ikx}dk=\frac{1}{
2\pi }\int k\psi (k)e^{-ikx}dk.
\end{equation}
Therefore,
\begin{equation}
(\rho {\widehat A}\omega \psi )(k)=k\psi (k).
\end{equation}
So, the eigenvalue problem in new coordinates is as follows:
\begin{equation}
\label{eigenvalue2}
g(k\psi )=p g(\psi ).
\end{equation}
Thus, we have the eigenvalue problem for the operator of multiplication by
the variable. The eigenfunctions here are given by
\begin{equation}
g(k)=\delta (p-k).
\end{equation}
Notice that $g=\omega ^{\ast }f$ is as it should be. Indeed,
\begin{equation}
(\omega ^{\ast }f)(k)=\frac{1}{2\pi }\int f(x)e^{-ikx}dx=\frac{1}{2\pi }\int
e^{ip x}e^{-ikx}dx=\delta (p-k ).
\end{equation}
As a result, the eigenvalue problems Eqs. (\ref{eigenvalue1}), and (\ref
{eigenvalue2}) can be considered as two coordinate expressions of a single string tensor equation Eq. (\ref{LLLL}).
 

Let us discuss now the differential geometry of string manifolds.
Assume that the string manifold under consideration is the abstract Hilbert space $\bf {S}$ itself. Choose a linear functional coordinate system $e^{-1}_{H}:{\bf S} \longrightarrow H$ on $\bf {S}$. 
Let $\Phi_{0}$ be a point in $\bf {S}$ and let $\Phi_{t}: R \longrightarrow {\bf S}$ be a differentiable path in $\bf {S}$ which passes through the point $\Phi_{0}$ at $t=0$.
Let $\varphi_{t}=e_{H}^{-1}(\Phi_{t})$ be the equation of the path in the basis $e_{H}$.

The vector $X$ {\it tangent} to the path $\Phi_{t}$ at the point $\Phi_{0}$ can be defined 
as the velocity vector of the path. In the basis $e_{H}$, $X$ is given by
\begin{equation}
\xi \equiv e^{-1}_{H}(X)=\left.\frac{d \varphi_{t}}{dt}\right|_{t=0}.
\end{equation}

Given vector $X$ tangent to $\Phi_{t}$ at the point $\Phi_{0}$ and a differentiable functional $F$ on a neighborhood of $\Phi_{0}$ in $\bf {S}$, the directional derivative of $F$ at $\Phi_{0}$ along $X$ is defined by
\begin{equation}
\label{tangent}
XF=\left.\frac{dF(\Phi_{t})}{dt}\right|_{t=0}.
\end{equation}
By applying the chain rule we have
\begin{equation}
\label{chain}
XF=
\left.F^{\prime}(\Phi)\right|_{\Phi=\Phi_{0}}\left.\Phi^{\prime}_{t}\right|_{t=0},
\end{equation}
where $F^{\prime}(\Phi)|_{\Phi=\Phi_{0}}: {\bf S} \longrightarrow R$ is the derivative functional at $\Phi=\Phi_{0}$ and $\Phi^{\prime}_{t}|_{t=0} \in {\bf S}$ is the derivative of $\Phi_{t}$ at $t=0$. Writing the last expression in coordinates yields
\begin{equation}
\label{var-d}
XF=
\int \left. \frac{\delta f(\varphi)}{\delta \varphi(x) }\right|_{\varphi=\varphi_{0}}\xi (x)dx,
\end{equation}
where $\xi=\varphi^{\prime}_{t}|_{t=0}$ and
$\left.\frac{\delta f(\varphi)}{\delta \varphi (x) }\right|_{\varphi=\varphi_{0}} \in H^{\ast}$, denotes the derivative functional $F^{\prime}(\Phi_{0})$ in the dual basis $e^{\ast}_{H}$. As before, the integral sign is understood here in the sense of action of $\frac{\delta f(\varphi)}{\delta \varphi (x) }$ on $\xi$.
In this notation we can also write symbolically
\begin{equation}
\label{var-d1}
X=\int \xi (x)\frac{\delta}{\delta \varphi(x)}dx.
\end{equation}
The right hand side of Eq. (\ref{var-d1}) acts on functionals $f$ defined by
\begin{equation}
f(\varphi)=F(\Phi),
\end{equation}
where $F$ is as before and $e_{H}\varphi=\Phi$. 

The space ${\bf T_{0}S}$ of all tangent vectors $X$ at a point $\Phi_{0}$ can be identified with the Hilbert space $\bf {S}$ itself and will be called the {\it tangent space} to $\bf {S}$ at the point $\Phi_{0}$. Notice also that the identification of ${\bf T_{0}S}$ with $\bf {S}$ makes it possible to identify the string basis $e_{H}$ with the local basis at $\Phi_{0}$ and with the symbol $\frac{\delta}{\delta \varphi}$.


Assume now that the kernel of the Hilbert metric on a coordinate space $H$ is a smooth function on $R^{n}\times R^{n}$. Then $H$ contains delta-functions and the subset $M$ of all delta-functions in $H$ forms a submanifold of $H$. In fact, it is easy to see that the map $a \longrightarrow \delta(x-a)$ is a smooth map from $R^{n}$ into $H$ which parametrizes the set $M$ of all delta-functions in $H$. Let us also remark that, although $M$ is not a linear subspace of $H$, any diffeomorphism $M\cong R^{n}$ induces a linear structure on $M$. In fact, if $\omega: R^{n} \longrightarrow M$ is a diffeomorphism, then we can define linear operations $\oplus,\odot$ on $M$ by $\omega(x+y)=\omega(x) \oplus \omega(y)$ and $\omega (kx)=k \odot \omega (x)$ for any vectors $x,y \in R^{n}$ and any number $k$. It is easy to check that these operations are continuos. The resulting linear structure on $M$ will be then different from the one on $H$.

In a similar way one can also derive topologically nontrivial spaces $M$. For example,
let $H$ be the Hilbert space of smooth functions on the interval $[0,2\pi]$ such that $\varphi^{(n)}(0)=\varphi^{(n)}(2\pi)$ for all $\varphi$ in $H$ and for all orders $n$ of (one-sided) derivatives of $\varphi$. Consider the dual space $H^{\ast}$ of functionals in $H$ and assume that the kernel of the metric on $H^{\ast}$ is smooth and that the space $H$ contains sufficiently many functions. Then the subset $M$ of delta-functions in $H$ form a submanifold diffeomorphic to the circle $S^{1}$ (see Ref. 9).

More generally, a Hilbert space $H$ of functions on an n-dimensional manifold can be identified with the space of functions on a subset of $R^{n}$. In fact, the manifold itself is a collection of non-intersecting ``pieces'' of $R^{n}$ ``glued'' together. Functions on the manifold can be then identified with functions defined on the disjoint union of all pieces and taking equal values at the points identified under ``gluing''. As a result, the dual space $H^{\ast}$ of generalized functions ``on'' the manifold can be also identified with the corresponding space of generalized functions ``on'' a subset of $R^{n}$. 

This fact allows us to conclude that topologically different manifolds $M$ can be obtained by choosing an appropriate Hilbert space of functions on a subset of $R^{n}$ and identifying $M$ with the submanifold of $H$ consisting of delta-functions. The manifold structure on $M$ is then induced by the embedding of $M$ into $H$ and does not have to be defined in advance. 

Moreover, the tangent bundle structure and the Riemannian structure on $M$ can be also induced by the embedding $i: M \longrightarrow H$.
To demonstrate this, let us select from all paths in $H$ the paths with values in $M$. In the chosen coordinates any such path $\varphi_{t}: [a,b] \longrightarrow M$  has the form
\begin{equation}
\label{path1}
\varphi_{t}(x)=\delta(x-a(t))
\end{equation}
for some function $a(t)$ taking values in $R^{n}$.


Vectors tangent to such paths can be identified with the ordinary $n$-vectors. In fact, assume $f$ is an analytic functional represented on a neighborhood of $\varphi_{0}=\left .\varphi_{t}\right |_{t=0}$ in $H$
by a convergent power series 
\begin{equation}
f(\varphi)=f_{0}+\int f_{1}(x)\varphi(x)dx+\int \int f_{2}(x,y)\varphi(x)\varphi(y)dxdy+...\quad ,
\end{equation}
where $f_{0}, f_{1}, f_{2}, ...$ are smooth functions.
Then on the path $\varphi_{t}(x)=\delta(x-a(t))$ we have
\begin{equation}
\label{4_vector}
\left.\frac{df(\varphi_{t})}{dt}\right|_{t=0}=
\left.\frac{\partial f(x)}{\partial x^{\mu}}\right|_{x=a(0)}\left.\frac{da^{\mu}}{dt}\right|_{t=0},
\end{equation}
where on a neighborhood of $a_{0}=a(0)$ in $R^{n}$ the function $f(a)=f(\delta_{a})$ with $\delta_{a}(x)= \delta(x-a)$ 
is given by the convergent series
\begin{equation}
f(a)=f_{0}+f_{1}(a)+f_{2}(a,a)+...\quad .
\end{equation}
In particular, the expression on the right of Eq. (\ref{4_vector}) can be immediately identified with the action of a n-vector $\frac{da^{\mu}}{dt} \frac{\partial}{\partial a^{\mu}}$  on the function $f(a)$. Using Eq. (\ref{var-d}) we also conclude that
\begin{equation}
\label{tT}
\int \left.\frac{d \varphi_{t}(x)}{dt}\right|_{t=0}\left. \frac{\delta f(\varphi)}{\delta \varphi(x)}\right |_{\varphi=\varphi_{0}}dx=\left. \frac{d a^{\mu}(t)}{dt}\right|_{t=0} \left.\frac{\partial f(a)}{\partial a^{\mu}}\right|_{a^{\mu}=a^{\mu}(0)}.
\end{equation}

Assume now that $H$ is a real Hilbert space and let $K: H \times H \longrightarrow R$ be the metric on $H$ given by a smooth kernel $k(x,y)$. 
If $\varphi=\varphi_{t}(x)=\delta(x-a(t))$ is a path in $M$, then for the vector $\delta \varphi(x)$ tangent to the path at $\varphi_{0}$ we have 
\begin{equation}
\delta \varphi(x)\equiv \left.\frac{d\varphi_{t}(x)}{dt}\right|_{t=0}=-\nabla_{\mu}\delta(x-a)\left.\frac {da^{\mu}}{dt}\right|_{t=0}. 
\end{equation}
Here $\nabla_{\mu}=\frac{\partial}{\partial x^{\mu}}$, $a=a(0)$ and derivatives are understood in a generalized sense, i.e. as linear functionals acting on smooth functions.  Therefore,
\begin{equation}
\left \| \delta \varphi \right \|^{2}_{H}=
\int k(x,y) \nabla{\mu}\delta(x-a)\left.\frac {da^{\mu}}{dt}\right|_{t=0}\nabla_{\nu}\delta(y-a)
\left.\frac {da^{\nu}}{dt}\right|_{t=0}dxdy.
\end{equation}
``Integration by parts" in the last expression gives
\begin{equation}
\int k(x,y) \delta \varphi(x)\delta \varphi(y)dxdy=
\left.\frac {\partial^{2}k(x,y)}{\partial x^{\mu} \partial y^{\nu}}\right|_{x=y=a} \left.\frac {da^{\mu}}{dt}\right|_{t=0}\left.\frac {da^{\nu}}{dt}\right|_{t=0}.
\end{equation}
By defining $\frac {da^{\mu}}{dt}|_{t=0}=da^{\mu}$, we have
\begin{equation}
\label{relation1}
\int k(x,y) \delta \varphi(x)\delta \varphi(y)dxdy=g_{\mu \nu}(a)da^{\mu}da^{\nu},
\end{equation}
where
\begin{equation}
\label{metric}
g_{\mu \nu}(a)=\left.\frac {\partial^{2}k(x,y)}{\partial x^{\mu} \partial y^{\nu}}\right|_{x=y=a}.
\end{equation}

As the functional $K$ is symmetric, the tensor $g_{\mu \nu}(a)$ can be assumed to be symmetric as well. If in addition $\left.\frac{\partial^{2}k(x,y)}{\partial x^{\mu} \partial y^{\nu}}\right|_{x=y=a}$ is positive definite at every $a$, the tensor $g_{\mu \nu}(a)$ can be identified with the Riemannian metric on an $n$-dimensional manifold $N$ diffeomorphic to $M$. 

In particular, consider the Hilbert space $H$ with metric given by the kernel $k({\bf x},{\bf y})=e^{-\frac{1}{2}({\bf x}-{\bf y})^{2}}$ for all ${\bf x},{\bf y} \in R^{3}$. Using Eq. (\ref{metric}) and assuming $({\bf x}-{\bf y})^{2}=\delta_{\mu\nu}(x^{\mu}-y^{\mu})(x^{\nu}-y^{\nu})$ with $\mu,\nu =1,2,3$, we immediately conclude that $g_{\mu\nu}(a)=\delta_{\mu\nu}$, which is the Euclidean metric.

The resulting isometric embedding is illustrated in Figure 1. The cones in the figure represent delta-functions forming the manifold $M$ which we denote in this case by $M_{3}$.
\begin{figure}[ht]
\label{fig:1}
\begin{center}
\includegraphics[width=8cm]{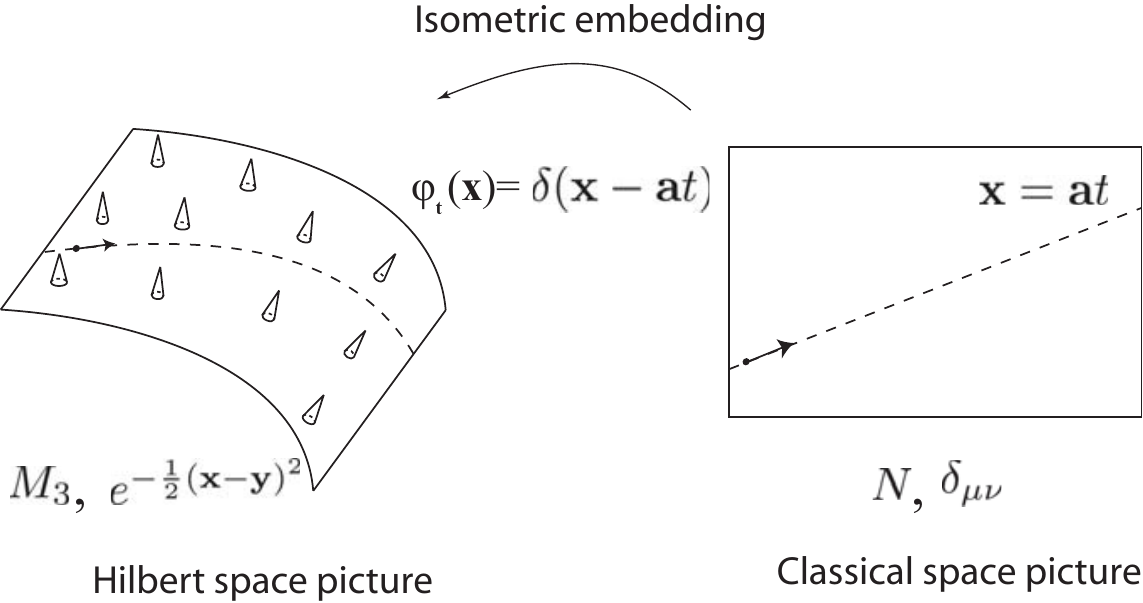}
\caption{\small{Isometric embedding of $R^{3}$ into $H$}}
\end{center}
\end{figure}

To understand better the embedding of $R^{3}$ into $H$ let us observe that the norm of any element $\delta({\bf x}-{\bf a})$ in $H$ is equal to $1$. Therefore, the three dimensional manifold $M_{3}$ is a submanifold of the unit sphere $S^{H}$ in $H$. 
Moreover, the set $M_{3}$ form a {\em complete system} in $H$. That is, there is no non-trivial element of $H$ orthogonal to every element of $M_{3}$. In fact, assume that $f$ is a functional in $H$ such that $\int e^{-\frac{1}{2}({\bf x}-{\bf y})^{2}}f({\bf x})\delta({\bf y}-{\bf u})d{\bf x}d{\bf y}=0$ for all ${\bf u} \in R^{3}$. Then $\int e^{-\frac{1}{2}({\bf x}-{\bf u})^{2}}f({\bf x})d{\bf x}=0$ for all ${\bf u} \in R^{3}$. 
Since the metric ${\widehat G}^{-1}: H^{\ast} \longrightarrow H$ given by the kernel $e^{-\frac{1}{2}({\bf x}-{\bf y})^{2}}$ is an isomorphism, we conclude that $f=0$.
It is also easy to see that the elements of any finite subset of $M_{3}$ are linearly independent. Indeed, if  $\sum_{k=1}^{n} c_{k}\delta(x-a_{k})$ is the zero functional in $H$ and the numbers $a_{k}$ are all different, then the coefficients $c_{k}$ must be all equal to zero. 
Finally, it is obvious that the set $M_{3}$ is uncountable and that no two elements of $M_{3}$ are orthogonal (although, provided $|a-b| \gg 1$, the elements $\delta(x-a)$, $\delta(x-b)$ are ``almost'' orthogonal).

The following two pictures help ``visualizing'' the embedding of $R^{3}$ into $H$. Under the embedding any straight line ${\bf x}={\bf a_{0}}+{\bf a}t$ in $R^{3}$ becomes a ``spiral'' $\varphi_{t}({\bf x})=\delta({\bf x}-{\bf a_{0}}-{\bf a}t)$ on the sphere $S^{H}$ through dimensions of $H$. One such spiral is shown in Figure 2. The curve in Figure 2 goes through the tips of three shown linearly independent unit vectors. Imagine that each point on the curve is the tip of a unit vector and that any $n$ of these vectors are linearly independent. 
\begin{figure}[ht]
\label{fig:2}
\begin{center}
\includegraphics[width=3cm]{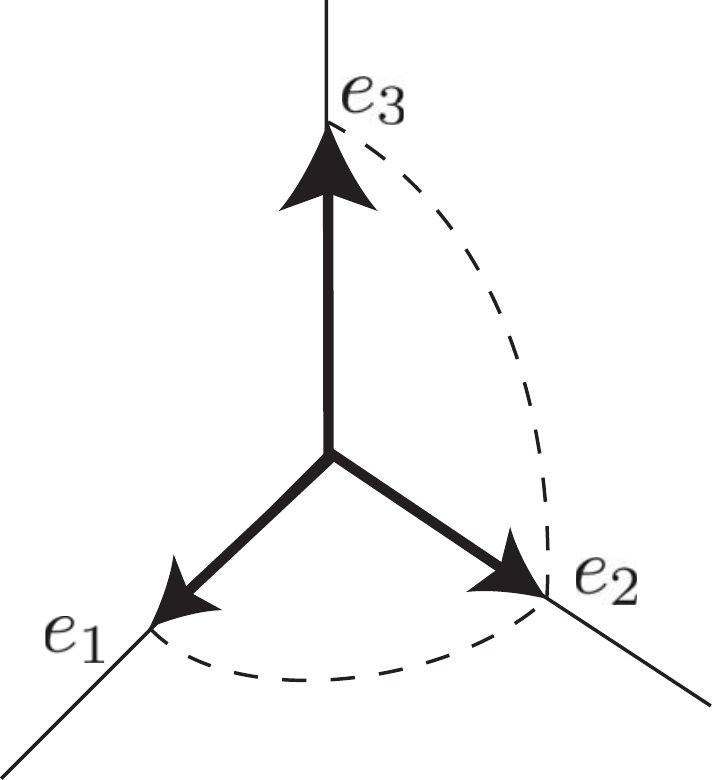}
\caption{\small{Straight line in $R^{3}$ as a ``spiral'' on the sphere $S^{H}$}}
\end{center}
\end{figure}

Based on this analysis, one can visualize the set $M_{3}$ as a three dimensional spiral-like submanifold in $S^{H}$ through the dimensions of $H$. Figure 3 illustrates the embedding of $R^{3}$ into $H$ in light of this result.
\begin{figure}[ht]
\label{fig:3}
\begin{center}
\includegraphics[width=5.5cm]{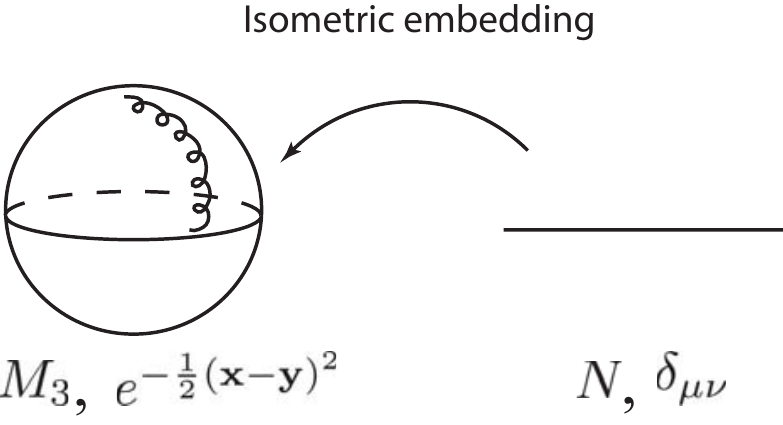}
\caption{\small{$R^{3}$ as a Riemannian submanifold of the sphere $S^{H}$}}
\end{center}
\end{figure}
Notice that under the embedding the infinite ``size'' of the Euclidean space $R^{3}$ has its counterpart in the infinite dimensionality of $S^{H}$.  

According to Ref. 8, {\em any} analytic Riemannian or pseudo-Riemannian metric on a finite dimensional manifold can be locally written in the form Eq. (\ref{metric}). In particular, for any analytic Riemannian or pseudo-Riemannian finite dimensional manifold $N$ there exists a coordinate Hilbert space $H$, such that $N$ is locally isometric to the submanifold $M$ of $H$ consisting of delta-functions. The described formalism will be referred to in the later sections as the {\em embedding formalism}.

\section{Observables as vector fields}

\setcounter{equation}{0}

Let us now assume that the classical space $M_{3}$ is embedded into a coordinate Hilbert space $H$ in the fashion described in Sec. 2. 
We saw that the Riemannian manifold structure on $M_{3}$ is induced in an elegant way by the embedding $i:M_{3} \longrightarrow H$. Our goal now is to reformulate QM in light of this embedding and to see to what extent such a reformulation may be useful.
The key observation is that the embedding $i:M_{3} \longrightarrow H$ allows one to extend the objects defined on the classical space to the entire Hilbert space. This extension will make the functional tensor approach to quantum theory possible.  

Consider for example the momentum operator $\widehat{p}^{\xi}=-i\xi^{\mu}\nabla_{\mu}$ ($\mu=1,2,3$) in the direction
specified by a unit vector $\xi$ in the classical space. 
By direct computation (and in agreement with Eq. (\ref{tT})), we have
\begin{equation}
\label{tT1}
\int \xi^{\mu} \nabla_{\mu}\delta(x-a) \left.\frac{\delta }{\delta \varphi(x)}\right |_{\varphi(x)=\delta(x-a)}dx=\xi^{\mu} \frac{\partial}{\partial a^{\mu}},
\end{equation}
where $\nabla_{\mu}=\frac{\partial}{\partial x^{\mu}}$, the left hand side acts on functionals of $\varphi$ and the right hand side acts on the corresponding functions on $R^{3}$. We conclude that, up to the factor $i$, the momentum operator $\widehat{p}^{\xi}$ is a restriction to the classical space $M_{3}$ of the linear in $\varphi$ string vector field $P_{\varphi}$ on $H$ defined by
\begin{equation}
\label{pP}
P_{\varphi}=-\int \xi^{\mu} \nabla_{\mu}\varphi(x) \frac{\delta }{\delta \varphi(x)}dx.
\end{equation}
Notice that because $M_{3}$ form a complete system in $H$,  the constructed linear extension $P_{\varphi}$ of the vector field Eq. (\ref{tT1}) from $M_{3}$ onto $H$ is unique.   

The above extension can be applied to any QM observable ${\widehat A}$ yielding a string vector field 
\begin{equation}
A_{\varphi}=e_{H}\left(-i{\widehat A}\varphi\right),
\end{equation} 
where the factor $-i$ has been used for the future convenience. In this case we will say that the vector field $A_{\varphi}$ is {\em associated} with the operator ${\widehat A}$.

In particular, the vector field associated with the position operator ${\widehat x}^{\eta}=\eta_{\mu}x^{\mu}$ in the direction of a unit covector $\eta$ is given by
\begin{equation}
Q_{\varphi}=-\int i\eta_{\mu}x^{\mu}\varphi(x)\frac{\delta}{\delta \varphi(x)}dx.
\end{equation}

For the commutator (Lie bracket) of vector fields $P_{\varphi}$ and $Q_{\varphi}$ we easily find:
\begin{equation}
\label{commutator1}
[P_{\varphi},Q_{\varphi}]=-i \int \eta_{\mu}\xi^{\mu}\varphi(x)\frac{\delta}{\delta \varphi(x)}dx.
\end{equation}
In particular, the commutator is again a vector field on $\bf {S}$ depending linearly on $\varphi$. 

More generally, assume that ${\widehat A}$, ${\widehat B}$ are observables and $A_{\varphi}$, $B_{\varphi}$ are the associated vector fields. Then one finds by a direct computation that
\begin{equation}
\label{commutator2}
[A_{\varphi},B_{\varphi}]=\int [{\widehat A},{\widehat B}]\varphi(x)\frac {\delta }{\delta \varphi(x)}dx,
\end{equation}
where $[{\widehat A},{\widehat B}]$ is the usual commutator of the observables. 

Given the vector field $A_{\varphi}$ associated with an observable ${\widehat A}$, consider an integral curve $\varphi_{\tau}$ of $A_{\varphi}$, i.e. the curve in $\bf {S}$ satisfying the equation
\begin{equation}
\label{integral}
\frac{d \varphi_{\tau}}{d \tau}=-i{\widehat A}\varphi_{\tau}.
\end{equation}
The general solution of Eq. (\ref{integral}) is given by
\begin{equation}
\label{curve}
\varphi_{\tau}(x)=e^{-i\tau {\widehat A}}\varphi_{0}(x),
\end{equation}
where $\varphi_{0}$ is the initial point on the curve. Indeed, since the observable ${\widehat A}$ is an Hermitian operator, Stone's theorem assures existence of the one-parameter group $e^{-i\tau {\widehat A}}$ of unitary operators with the generator $-i{\widehat A}$. 
Assume in particular that $\varphi_{0}$ is a unit-normalized state function in a Hilbert space $L_{2}$. Then the equation Eq. (\ref{curve}) describes a curve on the unit sphere $S^{L_{2}} \subset L_{2}$. 

Quite often the improper states can be approximated in some way by square integrable functions. Therefore the integral curves of observables passing through improper states can be still thought to be curves on the sphere $S^{L_{2}}$. Notice also that because delta-states can be approximated by the ``sharp'' Gaussian functions, the classical space can be identified in this approximation with a submanifold of $S^{L_{2}}$.

Alternatively, assume that $\varphi_{0}$ is an improper state that belongs to a Hilbert space $H$. For example, let $\varphi_{0}(x)=\delta(x-a)$ and let the space $H$ be defined by Eq. (\ref{example_H}). Then $\varphi_{0}$ does not belong to the sphere $S^{L_{2}}$ but is instead a point on the unit sphere $S^{H}$ in $H$ (recall that by Eq. (\ref{HHH}) the delta-function $\delta(x-a)$ is unit-normalized in $H$). Because the metrics on $H$ and $L_{2}$ are different, a transformation that is unitary transformation on $L_{2}$ is not necessarily unitary on $H$. As a result, the integral curves of observables are not guaranteed to take values in $S^{H}$. However, as discussed in Sec. 7 (see also Ref. 8), the metrics on $S^{L_{2}}$ and $S^{H}$ may be ``close'' to each other, so that the difference between the $L_{2}$ and the $H$-norm of a square-integrable function may not be significant. In this case the integral curves Eq. (\ref{curve}) through unit-normalized elements of either $L_{2}$ or $H$  can be considered  to be curves on the sphere $S^{H}$. At the same time the classical space $M_{3}$ is now a submanifold of $S^{H}$. 

However, the most appropriate way of working with several Hilbert metrics on a manifold at once is to consider the manifolds like $S^{L_{2}}$ and $S^{H}$ as Hilbert manifolds with a Riemannian metric $G$. The metric $G$ is then a tensor field which may vary along the manifold. In particular, the metric may be ``deformed'' along the submanifold $M_{3}$. The local coordinate charts may express this change in metric through the change in component functions of the metric and the corresponding change in the functional Hilbert space in which the charts take values. 

In the following, whenever the improper states are under discussion, the most convenient of the above three interpretations will be used. The notation $S^{G}$ will be used for the sphere $S^{L_{2}}$ furnished with a Riemannian metric $G$, i.e., for the pair $\left(S^{L_{2}}, G\right)$. Because any two separable infinite-dimensional Hilbert spaces are isomorphic, the spheres in these spaces are diffeomorphic. It follows that any Riemannian manifold diffeomorphic to a sphere in a Hilbert space can be identified with $\left(S^{L_{2}}, G\right)$ for some metric $G$. In particular, the unit sphere $S^{H}$ with Riemannian metric induced by embedding into $H$ can be identified with the sphere $S^{L_{2}}$ with a Riemannian metric $G$.

The vector field $A_{\varphi}={-i\widehat A}\varphi$ generates a motion of functionals along the integral curves $\varphi_{\tau}$. Namely,
if $f$ is a functional on $H$ and the values $\tau$, $\tau+\epsilon$ of the parameter mark the points $\varphi$ and $\varphi+\psi$ on an integral curve $\varphi_{\tau}$, then one can define a new functional $f_{\epsilon}$ by
\begin{equation}
f_{\epsilon}(\varphi_{\tau})=f(\varphi_{\tau+\epsilon}).
\end{equation}
Using the Taylor's series expansion we have
\begin{equation}
\label{Taylor}
f(\varphi_{\tau+\epsilon})=e^{\epsilon \frac{d}{d \tau}}f(\varphi_{\tau}).
\end{equation}
Alternatively, we can write
\begin{equation}
\label{Taylor1}
f(\varphi+\psi)=e^{\epsilon A_{\varphi}}f(\varphi)=e^{-i\int \epsilon {\widehat A}\varphi(x)\frac{\delta}{\delta \varphi(x)}dx}f(\varphi).
\end{equation}
According to Eq. (\ref{tT1}), for the vector field $P_{\varphi}$ associated with the momentum operator ${\widehat p}^{\xi}=-i\xi^{\mu}\nabla_{\mu}$, formula Eq. (\ref{Taylor1}) with terms restricted to $M_{3}$ reads
\begin{equation}
\label{Taylor3}
f(a+\epsilon \xi)=e^{\epsilon \xi^{\mu}\nabla_{\mu}}f(a).
\end{equation}
Here $f(a)$ is the value of the functional $f(\varphi)$ on delta-function $\delta_{a}(x)=\delta(x-a)$. A simple calculation shows that one could equivalently use the function ${\widetilde f}(a)=\left .\frac{\delta f(\varphi)}{\delta \varphi(x)}\right |_{\varphi(x)=\delta(x-a)}$ and replace the remaining variables $x$ with $a$ at the end. As follows from Eq. (\ref{Taylor3}), the Lie dragging of functions along vector fields on the classical space is a particular case of dragging functionals along string vector fields on the string space ${\bf S}$.

Let us now consider the integral curves of vector fields associated with momentum, energy and position observables in more detail.
From Eq. (\ref{curve}), we have for the momentum operator
\begin{equation}
\label{momentum}
\varphi_{\tau}(x)=e^{-\tau \xi^{\mu}\nabla_{\mu}}\varphi_{0}(x)=\varphi_{0}(x-\tau \xi),
\end{equation}
where the last equality is proved by a Taylor's series expansion. In particular, if $\varphi_{0}(x)=\delta(x-a)$, then
$\varphi_{\tau}(x)=\delta(x-a-\tau \xi)$. The resulting integral curve belongs in this case to the submanifold $M_{3} \subset S^{H}$ and the parameter $\tau$ can be identified with length in the classical space along the curve $\varphi_{\tau}$. 

For the energy operator ${\widehat h}=-\Delta+V(x)$ equation Eq. (\ref{integral}) is simply the Schr{\"o}dinger equation and we have
\begin{equation}
\label{energy}
\varphi_{\tau}(x)=e^{-i\tau {\widehat h}}\varphi_{0}(x).
\end{equation}
Accordingly, the parameter $\tau$ on the integral curve $\varphi_{\tau}$ in Eq. (\ref{energy}) is identified with time.

The integral curve of the vector field $Q_{\varphi}$ associated with the position operator is
\begin {equation}
\label{position}
\varphi_{\tau}(x)=e^{-i\tau \eta_{\mu}x^{\mu}}\varphi_{0}(x).
\end{equation}
To establish the meaning of parameter $\tau$ in this case let us apply the Fourier transform to $\varphi_{0}(x)$. From Eq. (\ref{position}) we obtain then
\begin{equation}
\label{position1}
\varphi_{\tau}(x)=e^{-i\tau \eta_{\mu}x^{\mu}}\int e^{ik_{\mu} x^{\mu}}\widetilde{\varphi}_{0}(k)dk=
\int e^{ip_{\mu} x^{\mu}}\widetilde{\varphi}_{0}(p+\tau \eta)dp,
\end{equation}
where $p=k-\tau \eta$.
That is, the Fourier image of $\varphi_{\tau}$ evolves by
\begin{equation}
\label{Fourier}
\widetilde{\varphi}_{\tau}(k)=\widetilde{\varphi}_{0}(k+\tau \eta).
\end{equation}
For simplicity, let us identify here the manifold $M_{3}$ with a submanifold of $S^{L_{2}}$ of sharp Gaussian functions which we still write in delta-function notation. 
Let us define the {\it momentum space} $\widetilde{M}_{3}$ to be the image of the space $M_{3}$ under the Fourier transform.
Since the Fourier transform is unitary in $L_{2}$, the momentum space is a submanifold of $S^{L_{2}}$. Clearly, the intersection $M_{3}\cap {\widetilde M}_{3}$ is empty. By Eq. (\ref{Fourier}) the integral curves of $Q_{\varphi}$ with $\varphi_{0}(k)=\delta(k-a)$ lie in ${\widetilde M}_{3}$ and  are given by $\varphi_{\tau}(k)=\delta(k+\tau \eta)$. Therefore, the parameter $\tau$ is the length along the curve $\varphi_{\tau}$ in the momentum space.

Note that the integral curves of the vector field $A_{\varphi}$ associated with ${\widehat A}$ form a {\em congruence}. That is, through each point $\varphi_{0} \in S^{L_{2}}$ such that ${\widehat A}\varphi_{0} \neq 0$  there passes a unique integral curve of $A_{\varphi}$ given by Eq. (\ref{curve}). This follows from the existence and uniqueness of the solution of Eq. (\ref{integral}) with the given initial state $\varphi_{0}$. 

Let us choose then a codimension one submanifold $\Omega \subset S^{L_{2}}$ of initial state functions transversal to the integral curves of $A_{\varphi}$ at least on a neighborhood $U \subset \Omega$ of a point $\varphi_{0}$. We can associate with each point $\varphi$ in a neighborhood $V$ of $\varphi_{0}$ in $S^{L_{2}}$ the pair $(\varphi_{0},\tau)$, $\varphi_{0} \in U$, $\tau \in R$, such that $\varphi=e^{-i{\widehat A}\tau}\varphi_{0}$. The pair $(\varphi_{0},\tau)$ can be used to parametrize $V$. We then call the above association a {\em partial one-dimensional coordinate system on $V$ associated with ${\widehat A}$} or simply the {\em ${\widehat A}$-coordinate system}.

Consider now two observables ${\widehat A}$ and ${\widehat B}$ and the corresponding vector fields $A_{\varphi}$ and $B_{\varphi}$. Suppose that the vector fields are linearly independent on a neighborhood of $\varphi_{0}$ in $L_{2}$  (and thus, by linearity of fields, on the entire $L_{2}$). Then the fields form what is called a {\em two-dimensional distribution} on $L_{2}$. By Frobenius theorem this distribution is integrable if and only if it is involutive. In other words, the integral curves of ${\widehat A}$ and ${\widehat B}$ ``sweep'' a family of two-dimensional submanifolds of $L_{2}$ if and only if the Lie bracket $[A_{\varphi},B_{\varphi}]$ is a linear combination of $A_{\varphi}$ and $B_{\varphi}$. 

In this situation let $\Omega \subset S^{L_{2}}$ be a codimension two submanifold of initial state functions which contains $\varphi_{0}$ and which is transversal to the integral curves of $A_{\varphi}$ and $B_{\varphi}$ at least on a neighborhood $U \subset \Omega$. Let $\tau, \lambda$ be parameters along the integral curves of $A_{\varphi}$ and $B_{\varphi}$ respectively. Then the triple $(\varphi_{0},\tau,\mu)$ can be used to parametrize a neighborhood of $\varphi_{0}$ in $S^{L_{2}}$ if and only if $[A_{\varphi},B_{\varphi}]=0$ on this neighborhood (equivalently, if and only if $[{\widehat A},{\widehat B}]=0$). In other words, the map
\begin{equation}
\label{rho}
\rho: (\varphi_{0},\tau,\lambda) \longrightarrow e^{-i{\widehat B}\lambda}e^{-i{\widehat A}\tau}\varphi_{0}
\end{equation}
from a neighborhood of $\varphi_{0}\times (0,0)$ in $U\times R^{2}$ into $S^{L_{2}}$ is a local diffeomorphism if and only if $[A_{\varphi},B_{\varphi}]=0$ (equivalently, if and only if $[{\widehat A},{\widehat B}]=0$). In this case we say that the pair $(U,\rho^{-1})$ is a {\em partial two-dimensional coordinate system on $V$} associated with operators ${\widehat A}$, ${\widehat B}$ or the {\em $\left\{{\widehat A}, {\widehat B}\right\}$-coordinate system}.

Figure 4 illustrates this result. The integral curves of $A_{\varphi}$, $B_{\varphi}$ in the figure  do not ``close up'' to form a coordinate grid unless $[{\widehat A},{\widehat B}]=0$.
\begin{figure}[ht]
\label{fig:4}
\begin{center}
\includegraphics[width=8cm]{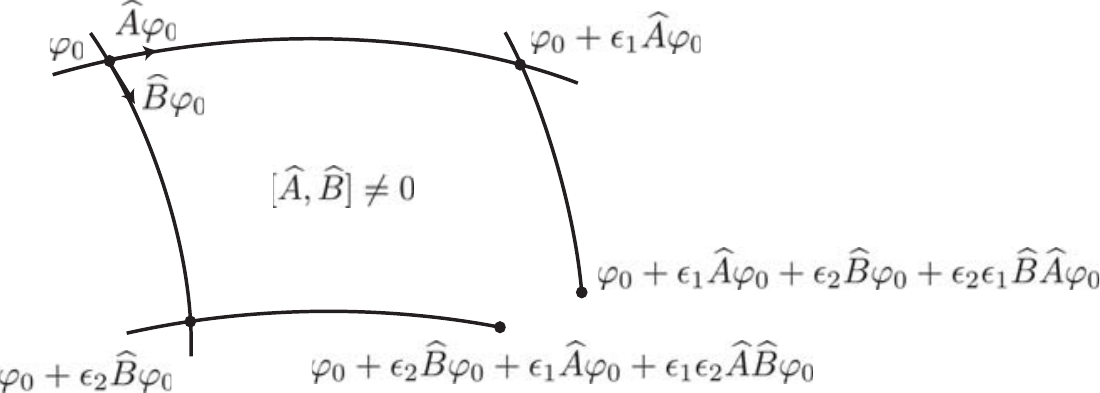}
\caption{\small{Integral curves of vector fields $A_{\varphi}$ and $B_{\varphi}$}}
\end{center}
\end{figure}

A similar analysis is valid for any finite number of observables and the associated vector fields.
We conclude that only when the observables under consideration commute do the integral curves of the associated vector fields form coordinate grids with parameters along the curves as coordinates of points belonging to the integral manifolds of the corresponding distributions. In particular, since components of the momentum operator ${\bf {\widehat p}}=-i{\bf \nabla}$ commute, the integral curves of the associated vector field  through the points $\delta(x-a)$ form a coordinate grid on $M_{3}$. Similarly, the integral curves of the vector field associated with the position operator ${\widehat{\bf x}}$ form a coordinate grid on the momentum space ${\widetilde M}_{3}$.

\section{Riemannian metric on the unit sphere $L_{2}$ and on the projective space $CP^{L_{2}}$}

\setcounter{equation}{0}

In the previous section we discussed integral curves of vector fields associated with various observables. 
The goal of this section is to demonstrate that the integral curves of vector field associated with Hamiltonian of a closed quantum system (i.e. solutions of the Schr{\"o}dinger equation for the system) are geodesics in the appropriate Riemannian metric on the space of states of the system. More generally, we will see that the integral curves of vector field associated with {\it any} observable with a trivial kernel are geodesics in the appropriate Riemannian metric. This fact will be important in Sec. 6, where the functionally covariant approach to quantum theory will be discussed. In establishing this fact we will also develop an infinite dimensional version of the local coordinate formalism on Riemannian manifolds.
  
In this section the index notation introduced in Sec. 2 will be used extensively. Thus, a string-tensor $T$ or rank $(r,s)$ in the index notation will be written as $t^{a_{1}...a_{r}}_{b_{1}...b_{s}}$. 
Assume that ${\widehat K}: H \longrightarrow H^{\ast}$ defines an Hermitian inner product $K(\xi,\eta)=(\widehat{K}\xi,\eta)$ on a complex Hilbert space $H$ of compex-valued functions $\xi$. Let $H_{R}$ be the real Hilbert space which is the {\it realization} of $H$. That is, $H_{R}$ is the space of pairs of vectors $(Re \xi, 
Im \xi)$, $\xi \in H$, with multiplication by real numbers. Alternatively, we can think of $H_{R}$ as the space of pairs  $X=(\xi, {\overline \xi})$ with multiplication by real numbers. In what follows the notation $H_{R}$ will always refer to this latter realization.

Since the inner product on $H$ is Hermitian, it defines a real valued Hilbert metric on $H_{R}$ by
\begin{equation}
K_{R}(X,Y)=2ReK(\xi,\eta),
\end{equation}
for all $X=(\xi, {\overline \xi})$, $Y=(\eta, {\overline \eta})$ with $\xi, \eta \in H$.
We will also use the ``matrix" representation of the corresponding operator ${\widehat K}_{R}: H_{R} \longrightarrow H_{R}^{\ast}$:
\begin{equation}
\label{metricKR}
\widehat {K}_{R}=\left[ 
\begin{array}{cc}
0 & {\widehat K} \\ 
{\overline {\widehat K}} & 0
\end{array}
\right].
\end{equation}
In particular, we have
\begin{equation}
K_{R}(X,Y)=({\widehat K}_{R}X,Y)=[\xi,{\overline \xi}]
{\widehat K}_{R}
\left[ 
\begin{array}{c}
\eta \\ 
{\overline \eta}
\end{array}
\right]=2Re({\widehat K}\xi,\eta),
\end{equation}
where $\xi{\widehat K}{\overline \eta}$ stands for the inner product $({\widehat K}\xi,\eta)$ and ${\overline \xi}{\overline{\widehat K}} \eta$ stands for its conjugate.

Let us agree to use the capital Latin letters $A, B, C, ...$ as indices of tensors defined on direct products of copies of the real Hilbert space $H_{R}$ and its dual. The small Latin letters $a, b, c, ...$ and the corresponding overlined letters ${\overline a}, {\overline b}, {\overline c}, ... $ will be reserved for tensors defined on direct products of copies of the complex Hilbert space $H$, its conjugate, dual and dual conjugate. A single capital Latin index replaces a pair of lower Latin indices. For example, if $X \in H_{R}$, then $X^{A}=(X^{a}, X^{{\overline a}})$, with $X^{a}$ representing an element of $H$ and  $X^{{\overline a}}={\overline X^{a}}$.  

Consider now the tangent bundle over a complex string space $\bf {S}$ which we identify here with a Hilbert space $L_{2}$ of square-integrable functions.  
Let us identify all fibers of the tangent bundle over $L_{2}$ (i.e. all tangent spaces $T_{\varphi}L_{2}$, $\varphi \in L_{2}$) with the complex Hilbert space $H$ described above. 
Let us introduce an Hermitian $(0,2)$ tensor field $G$ on the space $L_{2}$ without the origin as follows:
\begin{equation}
\label{Riem}
G(\xi,\eta)=\frac{({\widehat K}\xi,\eta)}{(\varphi,\varphi)_{L_{2}}},
\end{equation}
for all $\xi$, $\eta$ in the tangent space $T_{\varphi}L_{2}$ and all points $\varphi \in L_{2\ast}$. Here $L_{2\ast}$ stands for the space $L_{2}$ without the origin. 

The corresponding (strong) Riemannian metric $G_{R}$ on $L_{2}$ is defined by
\begin{equation}
G_{R}(X,Y)=2ReG(\xi,\eta),
\end{equation} 
where as before $X=(\xi, {\overline \xi})$ and $Y=(\eta, {\overline \eta})$.
In the matrix notation of Eq. (\ref{metricKR}) we have for the operator ${\widehat G}_{R}: H_{R} \longrightarrow H_{R}^{\ast}$ defining the metric $G_{R}$:
\begin{equation}
\label{metricGR_1}
{\widehat G}_{R}=\left[ 
\begin{array}{cc}
0 & {\widehat G} \\ 
{\overline {\widehat G}} & 0
\end{array}
\right],
\end{equation}
where ${\widehat G}: H \longrightarrow H^{\ast}$ defines the metric $G$.

In our index notation the kernel of the operator $\widehat {G}$ will be denoted by $g_{a{\overline b}}$, so that
\begin{equation}
\label{index_metric}
g_{a{\overline b}}=\frac{k_{a{\overline b}}}{\left\|\varphi \right\|^{2}_{L_{2}}},
\end{equation}
where $k_{a {\overline b}}$ is the kernel of ${\widehat K}$. 
From Eq. (\ref{metricGR_1}) we have for the components $({\widehat G}_{R})_{AB}$ of the metric ${\widehat G}_{R}$:
\begin{equation}
({\widehat G}_{R})_{ab}=({\widehat G}_{R})_{{\overline a}{\overline b}}=0,
\end{equation}
and
\begin{equation}
({\widehat G}_{R})_{a{\overline b}}=g_{a{\overline b}}, \quad
({\widehat G}_{R})_{{\overline a}b}={\overline g}_{a{\overline b}}.
\end{equation}
For this reason and with the agreement that $g_{{\overline a}b}$ stands for ${\overline g}_{a{\overline b}}$ we can denote the kernel of ${\widehat G}_{R}$ by $g_{AB}$. For the inverse metric we have
\begin{equation}
\label{metricGR}
{\widehat G}^{-1}_{R}=\left[ 
\begin{array}{cc}
0 & {\overline {\widehat G}^{-1}} \\ 
{\widehat G}^{-1} & 0
\end{array}
\right].
\end{equation}
Let the notation $g^{{\overline a}b}$ stand for the kernel of the inverse operator ${\widehat G}^{-1}$ and let $g^{a{\overline b}}$ stand for its conjugate ${\overline g^{{\overline a}b}}$. Then
\begin{equation}
({\widehat G}_{R})^{ab}=({\widehat G}_{R})^{{\overline a}{\overline b}}=0,
\end{equation}
and
\begin{equation}
({\widehat G}_{R})^{{\overline a}b}=g^{{\overline a}b},
({\widehat G}_{R})^{a{\overline b}}={\overline g^{{\overline a}b}}.
\end{equation}
Accordingly, without danger of confusion we can denote the kernel of ${\widehat G}^{-1}_{R}$ by $g^{AB}$.

Having the Riemannian metric $G_{R}$ on $L_{2}$ we can define the compatible ({\it Riemannian}, or {\it Levi-Civita}) connection $\Gamma$ by
\begin{equation}
\label{Levi}
2G_{R}(\Gamma(X,Y),Z)=dG_{R}X(Y,Z)+dG_{R}Y(Z,X)-dG_{R}Z(X,Y),
\end{equation}
for all vector fields $X,Y,Z$ in $H_{R}$. Here, for example, the term $dG_{R}X(Y,Z)$ denotes the derivative of the inner product $G_{R}(Y,Z)$ evaluated on the vector field $X$. 
In the given realization of the tangent bundle, for any $\varphi \in L_{2}$ the connection $\Gamma$ is an element of the space $L(H_{R},H_{R};H_{R})$. 
The latter notation means that $\Gamma$ is an $H_{R}$-valued $2$-form on $H_{R}\times H_{R}$.  
In our index notation the equation Eq. (\ref{Levi}) can be written as
\begin{equation}
2g_{AB}\Gamma^{B}_{CD}=\frac{\delta g_{AD}}{\delta \varphi^{C}}+\frac{\delta g_{CA}}{\delta \varphi^{D}}-\frac{\delta g_{CD}}{\delta \varphi^{A}}.
\end{equation}
Here for any $\varphi \in L_{2}$ the expression $g_{AB}\Gamma^{B}_{CD}$ is an element of $L(H_{R},H_{R},H_{R}; R)$, i.e., it is an $R$-valued $3$-form defined by
\begin{equation}
g_{AB}\Gamma^{B}_{CD}X^{C}Y^{D}Z^{A}=G_{R}(\Gamma(X,Y),Z)
\end{equation}
for all $X,Y,Z \in H_{R}$. Similarly, for any $\varphi \in L_{2}$, the variational derivative $\frac{\delta g_{AD}}{\delta \varphi^{C}}$ is an element of $L(H_{R},H_{R},H_{R}; R)$ defined by 
\begin{equation}
\frac{\delta g_{AD}}{\delta \varphi^{C}}X^{C}Y^{D}Z^{A}=dG_{R}X(Y,Z).
\end{equation}

For any $\varphi \in L_{2}$, by leaving vector $Z$ out, we can treat both sides of Eq. (\ref{Levi}) as elements of $H^{\ast}$.
Recall now that $G_{R}$ is a strong Riemannian metric. That is, for any $\varphi \in L_{2}$ the operator ${\widehat G}_{R}:H_{R} \longrightarrow H^{\ast}_{R}$ is an isomorphism, i.e., ${\widehat G}^{-1}_{R}$ exists. 
By applying ${\widehat G}^{-1}_{R}$ to both sides of Eq. (\ref{Levi}) without $Z$  we have in the index notation:
\begin{equation}
\label{Christoffel}
2\Gamma^{B}_{CD}=g^{BA}\left ( \frac{\delta g_{AD}}{\delta \varphi^{C}}+\frac{\delta g_{CA}}{\delta \varphi^{D}}-\frac{\delta g_{CD}}{\delta \varphi^{A}}\right ),
\end{equation}
where 
\begin{equation}
\Gamma^{B}_{CD}X^{C}Y^{D}\Omega_{B}=({\widehat G}^{-1}_{R}({\widehat G}_{R}\Gamma(X,Y)), \Omega).
\end{equation}
Formula Eq. (\ref{Christoffel}) defines the connection ``coefficients'' (Christoffel symbols) of the Levi-Civita connection. From the matrix form of ${\widehat G}_{R}$ and ${\widehat G}^{-1}_{R}$ we can now easily obtain
\begin{equation}
\label{11}
\Gamma^{b}_{cd}={\overline \Gamma}^{{\overline b}}_{{\overline c}{\overline d}}=\frac{1}{2}g^{{\overline a}b}\left (\frac {\delta g_{d{\overline a}}}{\delta \varphi^{c}}+\frac {\delta g_{c{\overline a}}}{\delta \varphi^{d}}\right ),
\end{equation}
\begin{equation}
\label{22}
\Gamma^{b}_{c{\overline d}}={\overline \Gamma}^{{\overline b}}_{{\overline c}d}=\frac{1}{2}g^{{\overline a}b}\left (\frac {\delta g_{c{\overline a}}}{\delta {\overline \varphi} ^{d}}-\frac {\delta g_{c{\overline d}}}{\delta {\overline \varphi}^{a}}\right),
\end{equation}
\begin{equation}
\label{33}
\Gamma^{b}_{{\overline c}d}={\overline \Gamma}^{{\overline b}}_{c{\overline d}}=\frac{1}{2}g^{{\overline a}b}\left (\frac {\delta g_{d{\overline a}}}{\delta {\overline \varphi} ^{c}}-\frac {\delta g_{{\overline c}d}}{\delta {\overline \varphi}^{a}}\right),
\end{equation}
while the remaining components vanish. To compute the coefficients, let us write the metric Eq. (\ref{index_metric}) in the form
\begin{equation}
g_{a{\overline b}}=\frac{k_{a{\overline b}}}{\delta_{u{\overline v}}\varphi^{u} {\overline \varphi}^{v}},
\end{equation}
where $\delta_{u{\overline v}}\equiv \delta(u-v)$ is the $L_{2}$ metric in the index notation.
We then have for the derivatives:
\begin{equation}
\frac{\delta g_{a{\overline b}}}{\delta \varphi^{c}}=-\frac{k_{a{\overline b}}\delta_{c{\overline v}}{\overline \varphi}^{v}}{\left\|\varphi \right\|^{4}_{L_{2}}},
\end{equation}
and
\begin{equation}
\frac{\delta g_{a{\overline b}}}{\delta {\overline \varphi}^{c}}=-\frac{k_{a{\overline b}}\delta_{u{\overline c}}\varphi^{u}}{\left\|\varphi \right\|^{4}_{L_{2}}}.
\end{equation}
Using Eqs. (\ref{11})-(\ref{33}) we can now find the non-vanishing connection coefficients
\begin{equation}
\label{111}
\Gamma^{b}_{cd}={\overline \Gamma}^{{\overline b}}_{{\overline c}{\overline d}}=-\frac{\left (\delta^{b}_{d}\delta_{c{\overline v}}+\delta^{b}_{c}\delta_{d{\overline v}}\right ){\overline \varphi}^{v}}{2\left\|\varphi \right\|^{2}_{L_{2}}},
\end{equation}
\begin{equation}
\label{222}
\Gamma^{b}_{c{\overline d}}={\overline \Gamma}^{{\overline b}}_{{\overline c}d}=-\frac{\left (\delta^{b}_{c}\delta_{u{\overline d}}-k^{{\overline a}b}k_{c{\overline d}}\delta_{u{\overline a}}\right )\varphi^{u}}{2\left\|\varphi \right\|^{2}_{L_{2}}},
\end{equation}
and
\begin{equation}
\label{333}
\Gamma^{b}_{{\overline c}d}={\overline \Gamma}^{{\overline b}}_{c{\overline d}}=-\frac{\left (\delta^{b}_{d}\delta_{u{\overline c}}-k^{{\overline a}b}k_{d{\overline c}}\delta_{u{\overline a}}\right )\varphi^{u}}{2\left\|\varphi \right\|^{2}_{L_{2}}}.
\end{equation}

Consider now the unit sphere $S^{L_{2}}: \left\|\varphi\right\|_{L_{2}}=1$ in the space $L_{2}$. Let ${\widehat A}$ be a (possibly unbounded) injective Hermitian operator defined on a set $D\left({\widehat A}\right)$ and with the image $R\left({\widehat A}\right)$. Here we assume for simplicity that $D\left({\widehat A}\right) \subset R\left({\widehat A}\right)$  and that both $D\left({\widehat A}\right)$ and $R\left({\widehat A}\right)$ are dense subsets of $L_{2}$. Let us define the inner product $(f,g)_{H}$ of any two elements $f,g$ in $R\left({\widehat A}\right)$ by the formula $\left(f,g\right)_{H} \equiv \left({\widehat A}^{-1}f, {\widehat A}^{-1}g\right)_{L_{2}}=\left (\left ({\widehat A}{\widehat A}^{\ast}\right)^{-1}f, g \right)$. By completing $R\left({\widehat A}\right)$ with respect to this inner product we obtain a Hilbert space $H$. Notice that ${\widehat A}$ is bounded in this norm and can be therefore extended to the entire space $L_{2}$. We will denote such an extension by the same symbol ${\widehat A}$. Let ${\widehat K}=({\widehat A}{\widehat A}^{\ast})^{-1}$, ${\widehat K}: H \longrightarrow H^{\ast}$ be the metric operator on $H$. As before, we define the Riemannian metric on $L_{2\ast}$ by
\begin{equation}
\label{RiemR}
G_{R}(X,Y)=\frac{2Re({\widehat K}\xi,\eta)}{(\varphi,\varphi)_{L_{2}}},
\end{equation}
where $X=(\xi,{\overline \xi})$, $Y=(\eta,{\overline \eta})$.
Assume that the sphere $S^{L_{2}} \subset L_{2\ast}$ is furnished with the induced Riemannian metric. Consider now the vector field $A_{\varphi}=-i{\widehat A}\varphi$ associated with the operator ${\widehat A}$. As in Sec. 3, the integral curves of this vector field are given by $\varphi_{\tau}=e^{-i{\widehat A}\tau}\varphi_{0}$. Since $e^{-i{\widehat A}\tau}$ denotes a  one-parameter group of unitary operators, the integral curve $\varphi_{\tau}$ through a point $\varphi_{0} \in S^{L_{2}}$ stays on $S^{L_{2}}$ . In particular, the vector field $A_{\varphi}$ is tangent to the sphere. In other words, the operator $-i{\widehat A}$ maps points on the sphere into vectors tangent to the sphere.

We claim now that the curves $\varphi_{\tau}=e^{-i{\widehat A}\tau}\varphi_{0}$ are geodesics on the sphere in the induced metric. That is, they satisfy the equation
\begin{equation}
\label{geophi}
\frac{d^{2}\varphi_{\tau}}{d\tau^{2}}+\Gamma\left (\frac{d\varphi_{\tau}}{d\tau},\frac{d\varphi_{\tau}}{d\tau}\right)=0.
\end{equation}
In fact, using Eqs. (\ref{111})-(\ref{333}) and collecting terms, we obtain
\begin{equation}
\label{geo1}
\Gamma^{b}_{CD}\frac{d\varphi^{C}_{\tau}}{d\tau}\frac{d\varphi^{D}_{\tau}}{d\tau}=\frac{\left({\widehat K}\frac{d\varphi_{\tau}}{d\tau},\frac{d\varphi_{\tau}}{d\tau}\right ){\widehat A}^{2}\varphi^{b}_{\tau}}{\left\| \varphi_{\tau}\right\|^{2}_{L_{2}}}.
\end{equation}
The expression for $\Gamma^{{\overline b}}_{CD}\frac{d\varphi^{C}_{\tau}}{d\tau}\frac{d\varphi^{D}_{\tau}}{d\tau}$ turns out to be the complex conjugate of Eq. (\ref{geo1}).
Now, the substitution of $\varphi_{\tau}=e^{i{\widehat A}\tau}\varphi_{0}$ and ${\widehat K}=\left ({\widehat A}{\widehat A}^{\ast}\right)^{-1}$ into the right hand side of Eq. (\ref{geo1}) yields ${\widehat A}^{2}\varphi_{\tau}$. At the same time, $\frac{d^{2}\varphi_{\tau}}{d\tau^{2}}=-{\widehat A}^{2}\varphi_{\tau}$ and therefore the equation Eq. (\ref{geophi}) is satisfied.
That is, the curves $\varphi_{\tau}=e^{-i{\widehat A}\tau}\varphi_{0}$ are geodesics in the metric Eq. (\ref{RiemR}) on $L_{2\ast}$. Since these curves also belong to the sphere $S^{L_{2}}$ and the Riemannian metric on the sphere is induced by the embedding $S^{L_{2}} \longrightarrow L_{2\ast}$, we conclude that the curves $\varphi_{\tau}$ are geodesics on $S^{L_{2}}$.
 
Assume in particular that ${\widehat A}$ is the Hamiltonian ${\widehat h}$ of a closed quantum system. Then the above model demonstrates that, in the appropriate Riemannian metric on the unit sphere $S^{L_{2}}$, the Schr{\"o}dinger evolution of the system is a motion along a geodesic of $S^{L_{2}}$. For a closely related metric on $S^{L_{2}}$ this result was obtained earlier in Ref. 8 by means of variational principle. 

Let us remark that the
formalism developed in this section is useful for other purposes as well. In particular, having the connection coefficients Eqs. (\ref{11})-(\ref{33}), we could have found the curvature of $S^{L_{2}}$ for the given Riemannian metric.

Notice also that multiplication by a non-zero complex number is an isometry of the metric Eq. (\ref{RiemR}). In other words, if $\lambda \in C_{\ast}$, where $C_{\ast}$ is the set of all non-zero complex numbers, then
\begin{equation}
G_{R}(\varphi)(X,Y)=G_{R}(\lambda \varphi)(d\lambda X, d\lambda Y).
\end{equation}
This follows at once from Eq. (\ref{RiemR}) and the fact that multiplication by a number is a linear map. We conclude that the metric Eq. (\ref{RiemR}) defines a Riemannian metric on the complex projective space $CP^{L_{2}}=L_{2\ast}/C_{\ast}$ of complex lines in $L_{2}$. When the space $H$ in Eq. (\ref{RiemR}) coincides with $L_{2}$, the resulting metric is nothing but the famous Fubini-Study metric on the infinite-dimensional space $CP^{L_{2}}$ (see Ref. 3). This metric will also show up in the finite dimensional setting that we are about to discuss.

\section{Riemannian metric in the 3-sphere $S^{3}$ and on the complex projective space $CP^{1}$}

\setcounter{equation}{0}

Instead of the infinite-dimensional sphere $S^{L_{2}}$ consider now the 3-sphere $S^{3}$ with the group structure of the Lie group $SU(2)$. The idea is to show that the formalism of the previous section has its natural counterpart in the Hilbert space $C^{2}$ of spin states of non-relativistic electrons. 
This puts us in the context of a well developed theory of Lie groups and homogeneous Riemannian manifolds. Accordingly, the exposition will be brief and the reader is referred to any standard text on the subject for details (for a simple practical approach, see Ref. 4).

Given an element ${\widehat A}$ of the Lie algebra $su(2)$, consider the left invariant vector field defined by $L_{{\widehat A}}(\varphi)=\varphi {\widehat A}$ for all $\varphi \in SU(2)$. The corresponding integral curve through a point $\varphi_{0} \in SU(2)$ has the form $\varphi_{\tau}=\varphi_{0}e^{{\widehat A}\tau}$. The {\em Killing metric} on $SU(2)$ can be defined by
\begin{equation}
\label{bb}
\left (L_{{\widehat A}}(\varphi), L_{{\widehat B}}(\varphi)\right )_{K}=-Tr \left(ad{\widehat A}\cdot ad{\widehat B}\right)
\end{equation}
for any ${\widehat A},{\widehat B} \in su(2)$.
Here the operator $ad{\widehat A}: su(2) \longrightarrow su(2)$ is defined by $ad{\widehat A}\left ({\widehat X} \right)=[{\widehat A},{\widehat X}]$ for all ${\widehat X} \in su(2)$ and similarly for $ad{\widehat B}$, and $Tr$ stands for the trace. Notice that the left invariant vector fields form a basis at any point $\varphi \in SU(2)$ and therefore the formula Eq. (\ref{bb}) defines the Riemannian metric on $SU(2)$. From the definition Eq. (\ref{bb}) we see that the Killing metric is invariant under the left and right action of $SU(2)$. Moreover, any other Riemannian metric with this property is proportional to the metric Eq. (\ref{bb}) and is also called the Killing metric.

Let us now define the connection $\nabla$ on $SU(2)$ by 
\begin{equation}
\label{aa}
\nabla_{L_{{\widehat A}}}L_{{\widehat B}}=\frac{1}{2}L_{[{\widehat A},{\widehat B}]}
\end{equation}
for any two left invariant vector fields. It is known that Eq. (\ref{aa}) defines the Levi-Civita connection of the Killing metric Eq. (\ref{bb}) (see Ref. 4). 
Moreover, the geodesics through identity element $e \in SU(2)$ are exactly the $1$-parameter subgroups of $SU(2)$. That is, for any ${\widehat A} \in su(2)$, the curve given by $\varphi_{\tau}=e^{{\widehat A}\tau}$ is the geodesic through $e$ in the direction of ${\widehat A}$. More generally, for any $\varphi_{0} \in SU(2)$ and any ${\widehat A} \in su(2)$ the integral curve $\varphi_{\tau}=\varphi_{0}e^{{\widehat A}\tau}$ of the vector field $L_{{\widehat A}}(\varphi)$ is the geodesic through $\varphi_{0}$ in the direction of ${\widehat A}$. 

We therefore see that, similarly to the infinite-dimensional case considered in the previous section, there exists a Riemannian metric on $S^{3}$ such that the integral curves of the linear vector field $\varphi {\widehat A}$ are geodesics on $S^{3}$. 

For the curvature tensor of the Killing metric $(\ , \ )_{K}$ on $SU(2)$ considered as a $(1,3)$-tensor evaluated on left invariant vector fields, we have
\begin{equation}
\label{curv1SU2}
R(L_{{\widehat A}},L_{{\widehat B}})L_{{\widehat C}}=-\frac{1}{4}L_{[[{\widehat A},{\widehat B}],{\widehat C}]}.
\end{equation}
When the curvature tensor is assumed to be a $(0,4)$-tensor, we have instead
\begin{equation}
\label{curv2SU2}
\left(R(L_{{\widehat A}},L_{{\widehat B}})L_{{\widehat C}},L_{{\widehat D}}\right)_{K}=\frac{1}{4}\left([{\widehat A},{\widehat B}],[{\widehat C},{\widehat D}]\right)_{K}.
\end{equation}
These formulas will be useful in Sec. 7.
 

The above formalism turns out to be relevant in physics. In fact,
the electron in the non-relativistic QM is described by a two-component state function. If one is only interested in the spin properties of the electron, its state function is a $C^{2}$-valued vector function of time. The values of this function are called {\em spin-vectors} or {\em spinors}. The sphere $S^{3}$ of unit spinors can be then identified with the group manifold $SU(2)$.

Since the states are physically determined only up to an overall phase factor, the physical space of states is  the projective space $CP^{1}=C^{2}_{\ast}/C_{\ast}$, where as before the asteric $\ast$ means ``take away zero".
The space $CP^{1}$ can be identified with the homogeneous space $SU(2)/S\left(U(1)\times U(1)\right)$. The group $SU(2)$ acts as a (transitive) group of transformations on $CP^{1}$ and $S\left(U(1)\times U(1)\right)$ can be identified with the isotropy subgroup mapping the circle $S^{1} \subset S^{3}$ representing the complex line through an arbitrary element $\varphi_{0} \in SU(2)$ into itself.

We can now decompose the Lie algebra $su(2)$ onto the orthogonal in the Killing metric sum of two subspaces $L^{0}$ and $L^{\bot}$.  Namely, the one-dimensional subspace $L^{0}$ is the Lie algebra of the isotropy subgroup of $\varphi_{0}$, while the two-dimensional subspace $L^{\bot}$ is the orthogonal complement of $L^{0}$. The space $CP^{1}$ can be then identified with the submanifold of $SU(2)$ spanned by geodesics through the identity element $e \in SU(2)$ in the direction of all vectors ${\widehat A} \in L^{\bot}$. As a result of this identification, the (positive definite) Killing metric on $SU(2)$ gives rise to the Riemannian metric on $CP^{1}$. In this Riemannian metric, $CP^{1}$ is a totally geodesic submanifold of $SU(2)$ and the integral curves $\varphi_{\tau}=\varphi_{0}e^{{\widehat A}\tau}$ of the vector fields $\varphi {\widehat A}$ with ${\widehat A} \in L^{\bot}$ are geodesics through $\varphi_{0}$ in the direction ${\widehat A}$.

The motion of a spinor $\varphi \in S^{3}=SU(2)$ along geodesic $\varphi_{\tau}=\varphi_{0} e^{{\widehat A}\tau}$ is projected by the bundle projection $\pi:C^{2}_{\ast} \longrightarrow CP^{1}$ to a motion on the base $CP^{1}$. 
The transformation properties of spinors under rotation admit a simple geometric interpretation in light of this projection. In essence, they are due to the fact that a plane (that is, a complex line, or a fibre) $C_{\ast}$ and the flipped upside down plane have the same image under the bundle projection $\pi$.

In particular, let us choose ${\widehat A}$ to be equal to $\frac{i}{2}{\widehat \sigma}_{3} \in su(2)$, where 
${\widehat \sigma}_{3}=
\left[ 
\begin{array}{cc}
1 & 0 \\ 
0 & -1
\end{array}
\right]$ 
is a Pauli matrix. 
Let 
\begin{equation}
\varphi_{\tau}=\varphi_{0}e^{\frac{i}{2}{\widehat \sigma}_{3}\tau}= 
\varphi_{0}\left[ 
\begin{array}{cc}
e^{\frac{i}{2}\tau} & 0 \\ 
0 & e^{-\frac{i}{2}\tau}
\end{array}
\right] 
\end{equation}
be the integral curve of the vector field $\varphi{\widehat A}$ through the spinor
$\varphi_{0}=\left[ 
\begin{array}{cc}
\xi & \eta \\ 
\end{array}
\right] \in S^{3}$.   
As we know, $\varphi_{\tau}$ is the geodesic through $\varphi_{0}$ in the direction $\varphi_{0}{\widehat A}$  in the Killing metric on $S^{3}$.
Under the motion along the geodesic the spinor $\varphi_{0}$ is transformed by
\begin{equation}
\left[ 
\begin{array}{cc}
\xi & \eta \\ 
\end{array}
\right]
\longrightarrow
\left[ 
\begin{array}{cc}
e^{\frac{i}{2}\tau}\xi & e^{-\frac{i}{2}\tau}\eta \\
\end{array}
\right].
\end{equation}
At the same time the complex line through 
$\left[ 
\begin{array}{cc}
\xi & \eta \\ 
\end{array}
\right]$
, which we denote by 
$\left\{ 
\begin{array}{cc}
\xi & \eta \\ 
\end{array}
\right\}$,
is transformed by 
\begin{equation}
\left\{ 
\begin{array}{cc}
\xi & \eta \\ 
\end{array}
\right\}
\longrightarrow
\left\{ 
\begin{array}{cc}
e^{\frac{i}{2}\tau}\xi & e^{-\frac{i}{2}\tau}\eta \\
\end{array}
\right\}=
\left\{ 
\begin{array}{cc}
e^{i\tau}\xi & \eta \\
\end{array}
\right\}.
\end{equation}
As $\tau$ changes from $0$ to $2\pi$, the spinor $\varphi_{\tau}$ changes from $\varphi_{0}$ to $-\varphi_{0}$, making half a revolution in $C^{2}$. At the same time, the plane $\pi\left (\varphi_{\tau}\right )=\left\{\varphi_{\tau}\right\}$, which for each $\tau$ is a point of $CP^{1}$, changes from 
$
\left\{ 
\begin{array}{cc}
\xi & \eta \\
\end{array}
\right\}
$ to 
$\left\{ 
\begin{array}{cc}
e^{i2\pi}\xi & \eta \\
\end{array}
\right\}$,
describing a full revolution about the $z$-axis in $R^{3}$ around the $2$-sphere $S^{2}$ identified with $CP^{1}$ (see Ref. 10). This is so because the spinors $\varphi_{0}$ and $-\varphi_{0}$ generate the same complex line $\left\{\varphi_{0}\right\}$.

Notice that if $\varphi_{0}$ is an eigenstate of ${\widehat \sigma}_{3}$, then the rotation is due to the phase factor only. In this case the corresponding path on $CP^{1}$ is trivial (i.e. the underlying point on $CP^{1}=S^{2}$ does not move). 

We remark here that the above projection of motion along $S^{3}$ onto a motion along $CP^{1}$ admits a very simple, almost mechanical interpretation described in Ref. 10.
It is also shown there that a similar interpretation of transformation properties of Dirac $4$-spinors describing relativistic electrons is valid.

Let us point out that the discussed Killing metric on $CP^{1}$ is proportional to the finite dimensional version of the previously mentioned Fubini-Study metric. Indeed, we could have derived both the Killing metric on $SU(2)$ and the corresponding metric on $CP^{1}$ by closely mimicking our derivation in the previous section.

In particular, we can identify the space $C^{2}$ of spinors with a subspace in a Hilbert space $L_{2}$ of $C^{2}$-valued state functions with the induced metric. Then the sphere $S^{3}$ of unit normalized spinors and the projective space $CP^{1}$ of physical spinors can be assumed to be isometrically and totally geodesically embedded submanifolds of the unit sphere $S^{L_{2}}$ and of the infinite-dimensional projective space $CP^{L_{2}}$ respectively. 
This embedding will be useful in Sec. 7.

\section{The principle of functional relativity}

\setcounter{equation}{0}

Physical reality in QT is independent of a particular representation used to describe it. In particular, when we transform an equation of motion in QT from the position to the momentum representation, the new equation describes the same underlying physical reality. At the same time the functional form of the equations of quantum theory in different representations is different. Consider for example the Klein-Gordon equation
\begin{equation}
\label{K-G}
\left(\partial_{\mu}\partial^{\mu}+\frac{m^{2}c^{2}}{\hbar^{2}}\right)\varphi(x)=0,
\end{equation}
which is a tensor equation under transformations of the Poincar{\'e} group $\Pi$. Note that here, in order to make the discussion more obvious, we will use a generic system of units and write all constants explicitly. When written in the momentum representation the equation Eq. (\ref{K-G}) becomes
\begin{equation}
\label{K-Gp}
\left(p_{\mu}p^{\mu}-m^{2}c^{2}\right)\psi(p)=0,
\end{equation}
which is a {\it different} tensor equation under the action of $\Pi$. In other words, the equations of QT considered as tensor equations on a group of space-time symmetry are not in general invariant under a change of representation.

Notice, however, that the {\it string} tensor form of the Klein-Gordon equation Eq. (\ref{K-G}) did not change. In fact, the equation can be written in an invariant way as 
\begin{equation}
\label{K-Gi}
\left({\bf {\widehat A}}_{\mu}{\bf {\widehat A}}^{\mu}-m^{2}c^{2}\right)\Phi=0.
\end{equation}
Here it is assumed that in a particular string basis $e_{H}$ the operator ${\bf {\widehat A}}_{\mu}$ is the operator of multiplication by the variable $p_{\mu}$:
\begin{equation}
\label{A_d}
e_{H}^{-1}{\bf {\widehat A}}_{\mu}e_{H}=p_{\mu}.
\end{equation}
In such a basis equation Eq. (\ref{K-Gi}) coincides with equation Eq. (\ref{K-Gp}). Then, in the Fourier transformed basis equation Eq. (\ref{K-Gi}) yields equation Eq. (\ref{K-G}).

In Sec. 2 we verified that the eigenvalue equations in QT can be also written in the string tensor form:
\begin{equation}
F\left({\bf {\widehat A}} \Phi \right)=\lambda F \left( \Phi \right).
\end{equation}
Moreover, in Sec. 3 the Schr{\"o}dinger equation was identified with the equation for integral curves of the vector field $-\frac{i}{\hbar}{\widehat h}\varphi$ associated with the Hamiltonian ${\widehat h}$:
\begin{equation}
\frac{d\varphi_{t}(x)}{dt}=-\frac{i}{\hbar}{\widehat h}\varphi_{t}(x).
\end{equation}
It is therefore a coordinate expression of a functional tensor equation on the string space $\bf {S}$. 
More generally, we saw in the previous sections that the main objects of QT can be all cast in a form that is independent of any particular functional realization. Examples include: quantum states $\Phi, \Psi,...$, the string space ${\bf S}$ to which these states belong, quantum observables ${\bf {\widehat A}},{\bf {\widehat B}},...$, vector fields ${\bf {\widehat A}}\Phi, {\bf {\widehat B}}\Phi,...$ associated with them, commutators of observables and of the associated vector fields, the previously mentioned eigenvalue problems and the Schr{\"o}dinger equation, etc. 

These results suggest that the quantum theory is a {\em functional tensor theory}. In other words,
\begin{flushleft}
\textsl{The laws of QT can be expressed in the form of functional tensor equations.}
\end{flushleft}
This hypothesis will be referred to as the {\it principle of functional relativity}. 
By itself the principle can be considered as simply a curious mathematical property of equations of QT. In fact, the transformations discussed so far in this section consisted in changing a particular functional realization $H$ needed to describe a physical reality without changing the string space ${\bf S}$ itself. Such transformations will be called {\it passive} as they are identity transformations on $\bf {S}$ being simply transformations of the sting basis $e_{H}$ on $\bf {S}$. To make the above principle of functional relativity into a physical principle, one must be able to realize the above transformations {\em physically}. To put it differently, one must be able to ``undo'' {\em any} passive transformation by the corresponding {\em active} transformation on ${\bf S}$.

 The situation is identical to the one in Galileo's thought experiment with the ship (see Ref. 2). The Galileo's principle of relativity is physical only because one can physically ``enclose yourself'' in the ship, observe various ``particulars'' and then ``make the ship move'', in which case ``You will not be able to discern the least alteration in all the ... effects'' (Ref. 2). In other words, there exists a physical transformation moving the entire Earth related laboratory to the ship in a uniform motion. This transformation is an {\em active} transformation in space complemented by (and ``compensated'' by) a Galilean transformation of the frame of reference.   

In the new setting the existence of active transformations in the string space ${\bf S}$ 
is immediately verified by any unitary evolution in QM. In this case ${\bf S}$ is identified with an $L_{2}$ space of state functions, and a unitary evolution operator is an automorphism of $L_{2}$. The Fourier transform experiment of Ref. 6 provides an example of evolution that is realized by an isomorphism of two different Hilbert spaces of functions. Since this experiment plays an important role in the coming discussion, let us briefly review it here.

A free electron from a source passes through a magnetic spectrometer and hits a vertical absorbing scintillating screen as shown on Figure 5.
\begin{figure}[ht]
\label{fig:5}
\begin{center}
\includegraphics[width=8cm]{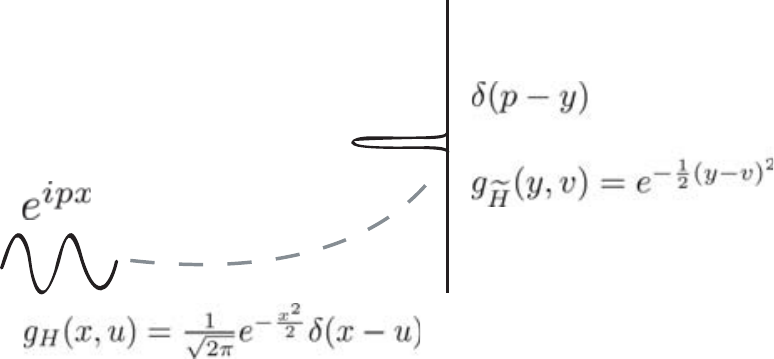}
\caption{\small{A thought experiment with magnetic spectrometer}}
\end{center}
\end{figure}
Due to the Lorentz force the electron will move in a circle of radius $r=\frac{p}{eB}$ (we neglect the effects related to spin and to emission of photons). Here $e$ is the electron's charge, $p$ is the magnitudes of electron's momentum $\bf{p}$, $B$ is the magnitude of the magnetic field $\bf{B}$, and the vectors $\bf {p}$ and $\bf {B}$ are assumed to be orthogonal. We conclude that position $y$ of the electron at the moment of absorption (see the figure) is uniquely determined by $p$.

Long enough before the electron enters the spectrometer, its wave function is an eigenstate of the momentum operator, i.e. it is proportional to $e^{ipx}$, where $x$ is the horizontal coordinate along the electron path. At the moment of absorption the state function of the electron can be assumed to be an eigenfunction of the position operator, i.e., it is proportional to $\delta(p-y)$. Here $y$ is the coordinate along the screen and the scale is chosen is such a way that the electron of momentum $p$ is absorbed at the point with $y=p$. 

We conclude that mathematically the spectrometer acts like the (inverse) Fourier transform:
\begin{equation}
F^{-1}\left[e^{ipx}\right](y)=\frac{1}{2\pi}\int e^{ipx}e^{-ixy}dx=\frac{1}{2\pi}\int e^{i(p-y)x}dx=\delta(p-y).
\end{equation}  
From the linearity of QM it follows that the spectrometer transforms superpositions of free electron states into superpositions of spatially localized electron states. The Hilbert space ${\widetilde H}$ of state functions of the electron which passed the spectrometer could be the space with the metric given by the kernel
\begin{equation}
\label{XX}
k_{{\widetilde H}}(y,v)=e^{-\frac{1}{2}(y-v)^{2}}.
\end{equation}
This metric was considered in Sec. 2 (we verified in Eq. (\ref{example_H}) that the corresponding Hilbert space contains delta-functions). The metric on the space $H$ is then the Fourier transformation of Eq. (\ref{XX}) by Eq. (\ref{tensor1}) and is given by the kernel
\begin{equation}
\label{moment}
k_{H}(x,u)=\frac{1}{\sqrt{2\pi}}e^{-\frac{x^{2}}{2}}\delta(x-u).
\end{equation}
The resulting space $H$ contains the free electron state functions of the initial electron.

The entire process can be described as an active transformation on ${\bf S}$ changing solutions of the generalized eigenvalue problem Eq. (\ref{eigenvalue1}) into the corresponding solutions of the generalized eigenvalue problem 
\begin{equation}
\label{eigenvalue3}
g(y\psi )=y g(\psi ).
\end{equation}
If the active Fourier transformation in the experiment is complemented by a change from coordinate to momentum representation, then the equation Eq. (\ref{eigenvalue3}) is changed back to 
\begin{equation}
\label{eigenvalue4}
f\left (-i\frac{d}{dp}\varphi \right )=x f(\varphi ).
\end{equation}

The above Fourier transform experiment followed by a change of representation mimics the Galileo's experiment with the ship. In fact, the physical transformation of state of an electron and of the observable in the experiment is ``compensated'' by the change of representation. As a result, the functional equations Eqs. (\ref{eigenvalue1}) and (\ref{eigenvalue4}) describing the electron before and after it passes through the spectrometer have the same form.

Let us demonstrate now that, in light of the embedding formalism of Sec. 2 (see also Ref. 3), the principle of functional relativity is a natural extension of the classical principle of relativity on space-time.
Let $N$ be the Minkowski space and let $\Lambda \in SO(1,3)$, $\Lambda:N \longrightarrow N$ be a Lorentz transformation acting on $N$. Assume that $H$ is a realization of $\bf {S}$ containing the submanifold $M_{4}$ of delta-functions identified with $N$ in the way described in Sec. 2. The kernel $\omega(x,y)=\delta(x-\Lambda y)$ defines a functional transformation $\omega$ on $H$ that maps $M_{4}$ into itself by
\begin{equation}
\label{Lorentz}
\int \delta(x-\Lambda y)\delta(y-a)dy=\delta(x-\Lambda a).
\end{equation}
We conclude that the transformation on the Minkowski space $N$ induced by the embedding $i: N \longrightarrow H$ maps $a \in N$ onto $\Lambda a$. In other words, the induced transformation is a Lorentz transformation. Moreover, the above transformations $\omega$ acting on $H$ form a group $L_{H}$ isomorphic to the Lorentz group $L=SO(1,3)$. In fact, if $\omega_{1}(x,y)=\delta(x-\Lambda_{1} y)$ and $\omega_{2}(x,y)=\delta(x-\Lambda_{2} y)$, then
\begin{equation}
\omega_{1}\omega_{2}(x,z)=\int \delta(x-\Lambda_{1}y)\delta(y-\Lambda_{2} z)dy=\delta(x-\Lambda_{1} \Lambda_{2} z).
\end{equation}
That is, the map defined by $\Lambda \longrightarrow \delta (x-\Lambda y)$ is an isomorphism of $L$ onto $L_{H}$. 

This result together with results of Sec. 2 can be summarized by saying that the tangent bundle over Minkowski space-time with the Lorentz group as a structure group is a subbundle of the tangent bundle over the string space. A similar statement holds true for more general tensor bundles. The covariance of tensor equations under Lorentz transformations is then induced by the above embedding. As a result, Einstein's principle of relativity is a special case of the principle of functional relativity.   

Moreover, the principle of functional relativity ascribes a new meaning to the speed of light $c$.
In fact, if $\varphi_{\tau}({\bf x})=\delta({\bf x}-{\bf a}(\tau))$ is a path with values in the space $M_{3} \subset H$ identified with the classical space $N$, then according to Eq. (\ref{relation1})
\begin{equation}
\label{c}
\left\|\frac{d\varphi_{\tau}}{d\tau}\right \|_{H}=\left\|\frac{d {\bf a}}{d\tau}\right \|_{N},
\end{equation}
where the metrics on $H$ and on $N$ are related by Eq. (\ref{metric}). Assume  that $N$ is the Euclidean 3-space $R^{3}$. Let $\tau$ be the classical time and let ${\bf a}(\tau)$ describe the motion of a classical particle. Then $\frac{d {\bf a}}{d\tau}$ is the velocity vector of the particle and the right hand side of Eq. (\ref{c}) cannot exceed the speed of light $c$. On the other hand, the left hand side of Eq. (\ref{c}) is a string-scalar, i.e. it is invariant under isomorphisms of Hilbert spaces. The immediate conclusion is that the speed of light is a string-scalar and not only a Lorentz scalar. 

In particular, since the motion of a classical particle is assumed to be physical, we expect it to be an approximation of the motion that satisfies the Schr{\"o}dinger equation with an appropriate Hamiltonian. Then, in accordance with the principle of functional relativity, any  coordinate transformation yields a physical equation of motion $\frac{d\psi_{\tau}}{d\tau}=-\frac{i}{\hbar}{\widehat A}\psi_{\tau}$ with the velocity $-\frac{i}{\hbar}{\widehat A}\psi_{\tau}$ of the norm less than $c$.
This observation will be important in application of the formalism to relativistic quantum theory.

The principle of functional relativity also leads one to an interesting conclusion about dimensions of observables in the theory. To see this, let us return to the Fourier transform experiment discussed earlier in this section. 
To make the discussion more obvious, let us use here the standard system of units. 
To simplify the expressions, let us assume that the vertical screen in Figure 5 goes through the centers of electron orbits so that the $y$ coordinate of the electron absorbed by the screen is given by $y=\frac{2p}{eB}$.
The kernels of the (active) Fourier transform and its inverse in the experiment are then given by
\begin{equation}
\label{F1}
\omega (x,y)=e^{-i\frac{xyeB}{2\hbar}}
\end{equation}
and
\begin{equation}
\label{F1-inv}
\omega^{-1}(y,x)=\frac{eB}{4\pi \hbar}e^{i\frac{yxeB}{2\hbar}}.
\end{equation}
Consider the equations for integral curves of vector fields associated with the position and momentum operators:
\begin{equation}
\label{pos}
\frac{d \varphi_{\tau}(x)}{d\tau}=-\frac{i}{\hbar}{\widehat x}\varphi_{\tau}(x)
\end{equation}
and
\begin{equation}
\label{mom}
\frac{d \psi_{\mu}(x)}{d\mu}=-\frac{i}{\hbar}{\widehat p}\psi_{\mu}(x).
\end{equation}
As already discussed, both Eqs. (\ref{pos}) and (\ref{mom}) are functional tensor equations expressed in functional coordinates. By applying the above active Fourier transform to both sides of Eq. (\ref{pos}), we obtain
\begin{equation}
\label{Four}
\frac{d \psi_{\tau}(y)}{d\tau}=-\frac{i}{\hbar}\frac{2}{eB}{\widehat p}\psi_{\tau}(y).
\end{equation}
Notice that the dimension of $eB$ is $\frac{P}{L}$, where $P$ is the dimension of momentum and $L$ is the dimension of length. For this reason the exponents in Eqs. (\ref{F1}), (\ref{F1-inv}) are dimensionless (as they should) and the terms on the left and the right hand sides of equations Eqs. (\ref{pos}) and (\ref{Four}) have the same dimension.

Let us now divide both sides of Eq. (\ref{Four}) by the coefficient $\frac{2}{eB}$:
\begin{equation}
\label{mom-1}
\frac{d \psi_{\tau}(y)}{d\left(\frac{eB}{2}\right)\tau}=-\frac{i}{\hbar}{\widehat p}\psi_{\tau}(y).
\end{equation}
Provided $\mu=\frac{eB\tau}{2}$ and $\psi_{\tau(\mu)}$ is identified with $\psi_{\mu}$, the equations Eqs. (\ref{mom-1}) and (\ref{mom}) can be now identified. In particular, since, as shown earlier, the dimension of $\tau$ in Eq. (\ref{pos}) is equal to $P$, the dimension of $\mu$ in Eq. (\ref{mom-1}) is $\frac{L}{P}\times P=L$.

There is an important lesson to be learned from this simple consideration. We know that there exists a coordinate transformation (change of representation) that relates the equations of integral curves of vector fields associated with operators of position and momentum. The principle of functional relativity insists then that such a transformation must be equivalent to the corresponding active transformation. 
The above example seems to be in agreement with this requirement. Notice however, that the active Fourier transform in the example needed to be complemented by division by the dimensional coefficient $\frac{2}{eB}$. In fact, we see from Eq. (\ref{Four}) that before the division the dimension of terms is not ``right''. 
The reason for that is clear: the position and momentum operators have different dimensions.
It follows that the functional principle of relativity can only be valid if dimensions of terms in the equations Eqs. (\ref{pos}) and (\ref{mom}) are equal. 

This conclusion can be clarified by an example in special relativity. For the special theory of relativity to be valid, the coordinates undergoing Lorentz transformation must have the same dimension. This is assured by introducing a new time variable $x^{0}=ct$ in place of the clock time $t$. Without this no ``mixing'' of space and time variables would be possible.

In the current case the operators ${\widehat x}$, ${\widehat p}$ at any point $\varphi_{0}$ on the sphere $S^{H}$ define two tangent directions $-i{\widehat x}\varphi_{0}$ and $-i{\widehat p}\varphi_{0}$. Accordingly,
the equations Eqs. (\ref{mom}) and (\ref{pos}) describe geodesics on $S^{H}$ through $\varphi_{0}$ in these two directions. Functional relativity requires ``mixing'' the directions. Therefore, the dimensions of terms ${\widehat x}\varphi$ and ${\widehat p}\varphi$ must be the same. This fact will be further clarified in the next section where we establish the functional-geometric nature of physical dimensions and of the commutators of observables.

\section{The origin of physical dimensions and of quantum commutators}

\setcounter{equation}{0}

Recall that the length of a line segment $[\varphi, \varphi+\delta\varphi]$ in a Hilbert space $H$ is given by
\begin{equation}
\label{length-1}
\left \|\delta\varphi\right\|^{2}_{H}=\int k(x,y)\delta\varphi(x)\delta\varphi(y)dxdy,
\end{equation}
where $k(x,y)$ is the kernel of the Hilbert metric on $H$. In the index notation of Sec. 2 this length can be written as
\begin{equation}
\label{length-2}
\left \|\delta\varphi\right\|^{2}_{H}=k_{xy}\delta\varphi^{x}\delta\varphi^{y}.
\end{equation}
The latter form of writing makes the meaning of the variables $x,y$ especially clear: they are just indices needed to label component functions of string tensors in a basis $e_{H}$. In particular, the equation Eq. (\ref{length-2}) is analogous to the equation $\left\|du\right\|^{2}=g_{\mu \nu}du^{\mu}du^{\nu}$ for the length element on a finite dimensional manifold with Riemannian metric $g$. 

As indices of tensor fields on a finite dimensional manifold carry no dimension, the indices $x,y$ in Eq. (\ref{length-2}) should be dimensionless as well. Moreover, the embedding formalism of Sec. 2 also supports the idea that the variables of functions $\varphi$ in a Hilbert space $H$ do not have a direct physical meaning. Instead, such a meaning is carried by the functions $\varphi$ themselves. Finally, according to the previous section, the principle of functional relativity can only be valid if dimensions of operators such as position and momentum coincide, in particular, if they are both dimensionless.

If the observables are indeed dimensionless, we must explain the way in which the standard interpretation of dimensions of physical quantities becomes possible. For this recall that in the embedding formalism of Sec. 2 the classical space $M_{3}$ is a submanifold of a Hilbert space $H$ formed by delta-functions. Moreover, the Riemannian metric on $M_{3}$ is induced by embedding via the formula
\begin{equation}
\label{relation}
\int k(x,y)\delta\varphi(x)\delta\varphi(y)dxdy=g_{\mu \nu}(a)da^{\mu}da^{\nu}.
\end{equation}
Here the metric $g_{\mu \nu}$ is given by Eq. (\ref{metric}). Assume now that the only dimensional quantities in the left hand side of Eq. (\ref{relation}) are functions $\varphi$ and that they carry the dimension of length $L$. Hence the left hand side of the equation Eq. (\ref{relation}) has dimension $L^{2}$ and the right hand side must have this dimension as well. In particular, in the case of the ordinary Euclidean metric $g_{\mu \nu}=\delta_{\mu \nu}$ we are forced to conclude that $da^{\mu}$ has dimension $L$. Therefore, the dimension of length on the classical space $M_{3}$ is induced via the embedding of $M_{3}$ into $H$. 

It is important to realize, however, that this method of inducing dimensions is not functionally covariant. In particular, as soon as we accept that the dimension of spatial coordinates $a^{\mu}$ is $L$, we are forced to recognize that the dimensions of momentum and position operators do not coincide. In particular, the operators ${\widehat p}$ and ${\widehat x}$ transform under a change of unit of length in a reciprocal way. 

So, the need for various physical dimensions may have its origin in the above identification of dimensions carried by functions and by the variables. The invariant approach to dimensions is to accept the dimension associated with functions as physical, consider the arguments of the functions as dimensionless and keep in mind that the right side of Eq. (\ref{relation}) is a special case of the functionally invariant expression on the left. 

With this accepted we need the length, time and momentum (or mass) to be dimensionless physical quantities. This by itself is easy to achieve by fixing an arbitrary system of units and considering dimensionless ratios (for example, length divided by the unit length, time divided by the unit time, etc). A similar ``cancellation'' of dimensions can be done in physical equations relating dimensional quantities. However, the ratios of length, time and mass will depend in this case on the chosen system of units. Because of that we need a system of units that would be physical, rather than ``anthropomorphic''. In other words, the units in such a system must be independent of any particular human convention.

Such a system of units is well known and, in fact, widely used in high energy physics. It is the so-called Planck system of units in which $c=\hbar=\gamma=1$ with $\gamma$ being the constant of gravity. The units of length, time and mass in this system (the Planck length $l_{P}$, time $t_{P}$ and mass $m_{P}$) can be expressed in terms of the standard SI units as follows:
\begin{eqnarray}
\label{lP}
l_{P}& \approx &1.6 \cdot 10^{-35}m,\\
t_{P}& \approx &5.4 \cdot 10^{-44}s, \\
m_{P}& \approx &2.2 \cdot 10^{-8}kg.
\end{eqnarray}
When physical quantities are expressed in Planck units they become dimensionless physically meaningful numbers (such as length divided by the Planck length, time divided by the Planck time, etc.) 
Since the Planck units are defined in terms of the physical constants $c, \hbar, \gamma$, they would change in any physical process that changed these physical constants. At the same time the values of physical quantities would change under these circumstances in a similar fashion. Because of that their expression in Planck units would remain unchanged provided the dimensionless physical constants stay the same (see Ref. 5). 

From now on we will assume that the values of physical quantities are always expressed in Planck units as dimensionless ratios. Then the position and momentum operators become dimensionless and have the form
\begin{eqnarray}
{\widehat x}&=&x,\\
{\widehat p}&=&-i\frac{d}{dx}.
\end{eqnarray}
The Fourier transform relates the two while preserving their dimensionlessness.
The equation for integral curves of the vector field associated with an observable ${\widehat A}$ in Planck units has a simple form
\begin{equation}
\label{inv-evolution}
\frac{d\varphi_{\tau}}{d\tau}=-i{\widehat A}\varphi_{\tau},
\end{equation}
where the operator ${\widehat A}$ and the parameter $\tau$ are dimensionless. 
The equation Eq. (\ref{inv-evolution}) has been already used earlier in the paper without much discussion.

Recall that according to Sec. 2 the Euclidean metric on the classical space $M_{3}$ can be induced by the embedding $i:M_{3} \longrightarrow H$, where $H$ is the Hilbert space with the metric $K$ given by the kernel $e^{-\frac{1}{2}(x-y)^{2}}$. We saw that the space $H$ contains delta-functions and that the expectation value of the position operator ${\widehat x}$ for a particle in state $\delta(x-a)$ is equal to $a$. It was also pointed out in Sec. 2 that not all of the results of the standard QM can be exactly reproduced in metric $K$. However, we are going to demonstrate now that within applicability of the standard QM, the difference between its predictions and the results of corresponding calculations in metric $K$ is too small to be detected in any current experiment.

For instance, an easy calculation demonstrates that the norm of superposition $c_{1}\delta(x-a)+c_{2}\delta(x-b)$ of two position eigenstates in metric $K$ is equal to 
\begin{equation}
\label{cccc}
|c_{1}|^{2}+|c_{2}|^{2}+\left(c_{1}{\overline c_{2}}+{\overline c_{1}}c_{2}\right)e^{-\frac{1}{2}(a-b)^{2}}.
\end{equation}
Recall now that the variables are measured here in Planck units. Also, the current experiments can only resolve distances significantly larger than the Planck length. Therefore, for superposition of any physically distinguishable position eigenstates the norm of $a-b$ in Planck units is a very large number. Therefore, the exponent $e^{-\frac{1}{2}(a-b)^{2}}$ is negligibly small and the equation Eq. (\ref{cccc}) reproduces the expected result with an extremely high accuracy. Clearly, the result can be easily generalized to arbitrary finite compositions of delta functions and to various bilinear expressions evaluated on such compositions.

Moreover, the results of calculations in metric $K$ are also extremely accurate for the system in an arbitrary square integrable state. For instance, consider a particle in a bound state $\varphi$ in one dimension and let us evaluate the norm of $\varphi$ in $K$ metric. This norm is given by 
\begin{equation}
\label{conti}
\left\|\varphi\right\|_{K}^{2}=\int e^{-\frac{1}{2}(x-y)^{2}}\varphi(x)\varphi(y)dxdy.
\end{equation}
As before, the variables $x$ and $y$ in Eq. (\ref{conti}) are measured in Planck units. Let us denote the length variable $x$ measured in macroscopic length units, say meters, by $x_{L}$. We then have $x=L x_{L}$, where according to Eq. (\ref{lP}) the coefficient $L$ is of the order of $10^{35}$. Using Eq. (\ref{conti}) and denoting $\varphi(Lx_{L})$ by $\psi(x_{L})$, we have
\begin{equation}
\label{conti_1}
\left\|\varphi\right\|_{K}^{2}=L\sqrt{\pi}\int \frac{L}{\sqrt \pi}e^{-\frac{1}{2}L^{2}(x_{L}-y_{L})^{2}}\psi(x_{L})\psi(y_{L})dx_{L}dy_{L}.
\end{equation}
It is known that the sequence $k_{L}(x_{L},y_{L})=\frac{L}{\sqrt \pi}e^{-\frac{1}{2}L^{2}(x_{L}-y_{L})^{2}}$ is a delta-convergent sequence as $L \longrightarrow \infty$. In other words, for large $L$ the kernel $k_{L}(x_{L},y_{L})$ behaves as the delta-function $\delta(x_{L}-y_{L})$. Since $L$ is of the order of $10^{35}$, we conclude that the value of the integral in Eq. (\ref{conti_1})
is extremely close to the standard expression $\left \|\psi\right \|_{L_{2}}^{2}$. The coefficient $L\sqrt{\pi}$ in front of the integral indicates that the expressions $\left\|\varphi\right\|_{H}$ and $\left\|\psi\right\|_{L_{2}}$ are normalized differently. This, however, does not affect the measurable predictions of quantum theory. Generalization of this result to various bilinear expressions is immediate.

The above metric $K$ evaluated in momentum representation yields the metric ${\widetilde K}$ with the kernel $\frac{1}{\sqrt{2\pi}}e^{-\frac{k^{2}}{2}}\delta(k-p)$. The fact that for the square integrable states the metric $K$ is practically indistinguishable from the $L_{2}$-metric has its natural counterpart in the case of metric ${\widetilde K}$. In fact, since the norm of momentum $k$ of a particle in the modern quantum mechanical experiments is much smaller than the Planck unit of mass (see Eq. (\ref{lP})), the exponent $e^{-\frac{k^{2}}{2}}$ can be safely replaced with $1$.

With these results in hand we are ready to investigate the meaning of commutators of observables in quantum theory. 
Let $L_{2}$ be a space of $C^{2}$-valued square-integrable functions and let $S^{G}$ be the unit sphere in $L_{2}$ with a Riemannian metric $G$ on it. 
Assume as in Sec. 5 that the sphere of unit spinors $S^{3}=SU(2)$ with the Killing metric is embedded isometrically and totally geodesically into $S^{G}$. Accordingly, the space of projective spinors $CP^{1}=S^{3}/S^{1}$ with the induced Fubini-Study metric is embedded isometrically and totally geodesically into the projective space $CP^{L_{2}}$ furnished with the Riemannian metric induced by embedding $CP^{L_{2}} \longrightarrow S^{G}$. 

The results of Secs. 4 and 5 suggest that there exists a Riemannian metric on $S^{G}$ in which the integral curves of the vector fields associated with observables of interest are geodesics. 
In the considered models this fact was verified for a single observable with a trivial kernel and for the spin observables. 

Assume then that ${\widehat A}$, ${\widehat B}$ are observables, and that $-i{\widehat A}\varphi$, $-i{\widehat B}\varphi$ are the corresponding vector fields and the integral curves $e^{-i{\widehat A}\tau}\varphi_{0}$, $e^{-i{\widehat B}\tau}\varphi_{0}$ are geodesics of $S^{G}$. Then the sectional curvature of $S^{G}$ in the plane through tangent vectors $-i{\widehat A}\varphi$, $-i{\widehat B}\varphi$ at any point $\varphi_{0}$ can be expressed in terms of the commutators of these fields. 

Suppose for example that ${\widehat A}$ and ${\widehat B}$ are spin observables.
Recall that in the Planck system of units the operator of spin ${\bf {\widehat s}}$ has eigenvalues $\pm 1/2$ and can be expressed in terms of the Pauli matrices ${\widehat \sigma}_{1}, {\widehat \sigma}_{2}, {\widehat \sigma}_{3}$ as
\begin{equation}
{\bf {\widehat s}}=\frac{1}{2}{\bf {\widehat \sigma}}
\end{equation}
with ${\bf {\widehat \sigma}}=({\widehat \sigma}_{1}, {\widehat \sigma}_{2}, {\widehat \sigma}_{3})$. 
The corresponding anti-Hermitian generators ${\widehat e}_{k}=\frac{i}{2} {\widehat \sigma}_{k}$ form a basis of the Lie algebra $su(2)$ and satisfy the commutator relations
\begin{equation}
\left[{\widehat e}_{k},{\widehat e}_{l}\right]= \epsilon _{klm}{\widehat e}_{m},
\end{equation}
where $\epsilon_{klm}$ denotes the completely antisymmetric tensor of rank three.

Recall now that any vector $x=(x^{k})$ in the Euclidean space $R^{3}$ can be identified with the element $ix^{k}{\widehat \sigma}_{k}=2x^{k}{\widehat e}_{k}$ of the Lie algebra $su(2)$. Then the Euclidean norm $\left\|x\right\|_{R^{3}}$ of $x$ is equal to $det (x)$ and rotations in $R^{3}$ are represented by transformations $x \longrightarrow UxU^{+}$ with $U \in SU(2)$. 

Let us accept this identification and let us also recall that the embedding of $R^{3}$ into $S^{G}$ is assumed to be isometric. 
Notice that the Killing metric on $S^{3} \subset S^{G}$ is defined up to a constant factor and in any Killing metric ${\widetilde K}$ on $S^{3}$ we have $\left(2x^{k}{\widehat e}_{k}, 2x^{m}{\widehat e}_{m}\right)_{{\widetilde K}}=4x^{k}x^{m}{\widetilde g}_{km}$, where ${\widetilde g}_{km}=\left({\widehat e}_{k},{\widehat e}_{m}\right)_{{\widetilde K}}$ are the components of ${\widetilde K}$ in the basis ${\widehat e}_{k}$. To satisfy the isometric embedding condition we must have then ${\widetilde g}_{km}=\frac{1}{4}\delta_{km}$.

At the same time, the components $g_{km}$ of the Killing metric Eq. (\ref{bb}) in the basis ${\widehat e}_{k}$ are given by
$g_{km}=2\delta_{km}$.
In other words, the Killing metric Eq. (\ref{bb}) must be multiplied by $\frac{1}{8}$. This also means that the corresponding sectional curvature of the Killing metric on $S^{3}=SU(2)$ will be multiplied by $8$.
Using the formula Eq. (\ref{curv2SU2}), we then have for the sectional curvature 
$R(p)$ in the plane $p$ through orthogonal vectors $L_{{\widehat e}_{1}},L_{{\widehat e}_{2}}$: 
\begin{equation}
\label{section}
8 \cdot  \frac{\left(R(L_{{\widehat e}_{1}},L_{{\widehat e}_{2}})L_{{\widehat e}_{2}},L_{{\widehat e}_{1}}\right)_{K}}{\left(L_{{\widehat e}_{1}},L_{{\widehat e}_{1}}\right)_{K}\left(L_{{\widehat e}_{2}},L_{{\widehat e}_{2}}\right)_{K}}=8 \cdot \frac{1}{4}\cdot \frac{ \left([{\widehat e_{1}},{\widehat e_{2}}],[{\widehat e_{1}},{\widehat e_{2}}]\right)_{K}}{4}=\frac{1}{2}\left ({\widehat e}_{3},{\widehat e}_{3}\right)_{K}=1.
\end{equation}
This sets the radius of $S^{3}$ in Planck units at $1$.

It follows that, at least in the directions specified by the spin observables, $S^{G}$ is an extremely small sphere. According to Eq. (\ref{lP}), it is about $10^{-35}$ of a meter in diameter. Despite the apparent minuscule size of the sphere $S^{G}$, the classical space can be isometrically embedded into it. In particular, we verified in Sec. 2 that the Euclidean space $R^{3}$ can be isometrically embedded into $S^{G}$ as a ``spiral'' through the dimensions of $S^{G}$. We also remark that the obtained radius of $S^{G}$ is exactly equal to the {\em minimal length} that is widely believed to exist in quantum gravity. In particular, the notion of minimal length acquires an unexpected geometric interpretation.

This picture reveals the dual role of Planck's constant. First of all, in a ``dimensionfull'' system of units such as SI, it plays the role of a dimensional coefficient needed to relate the dimensions of length $L$ and momentum $P$. In this respect $\hbar$ is similar to the speed of light $c$ relating the dimensions of length and time.

More importantly, the geometric meaning of $\hbar$ becomes clear when looking at the commutators of observables that contain $\hbar$. Namely, according to Eq. (\ref{section}) the commutators of observables are directly related to the sectional curvature of $S^{G}$. In other words, according to the theory, the non-trivial commutators of observables in QM are related to the non-vanishing curvature of the sphere $S^{G}$. At the same time the smallness of Planck's constant in SI units has its origin in the minuscule size of $S^{G}$ in these units.

\section{Application to the process of measurement}

\setcounter{equation}{0}

One of the most important consequences of the principle of functional relativity is that quantum processes (including quantum measurements) take place on an infinite-dimensional Hilbert manifold rather than on classical space. This observation turns out to be crucial in providing a strikingly simple interpretation of quantum mechanical experiments.  
For illustration let us consider the famous two-slit experiment with electrons. 

Assume that the function $\varphi_{\tau}=\varphi_{\tau}({\bf x})$ describes the initial wave packet of a free electron propagating toward the screen with the slits. Let us denote the Hamiltonian of the system by ${\widehat h}$ and let us identify the parameter $\tau$ with time.
As we know, the path $\varphi_{\tau}$ is a geodesic in the Riemannian metric ${\widehat G}=\left(hh^{\ast}\right)^{-1}$ on $S^{L_{2}}$. As in Sec. 3, in the ${\widehat h}$-coordinate system on a neighborhood of $\varphi_{0}=\left. \varphi_{\tau}\right|_{\tau=0}$ the path has a simple form, which is linear in $\tau$
\begin{equation}
\varphi_{\tau} =(\varphi_{0}, \tau).
\end{equation} 

Assume that $\chi_{\tau }$ and $\xi_{\tau }$ are (unit normalized) state functions of the electron that passed through one of the slits with the other slit closed. Then the state function of the electron that has passed through the screen with both slits open is a superposition
\begin{equation}
\psi_{\tau}=a\chi_{\tau }+b\xi_{\tau},
\end{equation}
where $a,b \in C$ and $|a|^{2}+|b|^{2}=1$.
The path $\psi_{\tau}$ is a geodesic in the metric ${\widehat G}$ and its equation in ${\widehat h}$-coordinates is
\begin{equation}
\psi_{\tau} =(\psi_{0}, \tau).
\end{equation} 
The entire process of passing through the slits expressed in ${\widehat h}$-coordinates is shown in Figure 6.
\begin{figure}[ht]
\label{fig:6}
\begin{center}
\includegraphics[width=4cm]{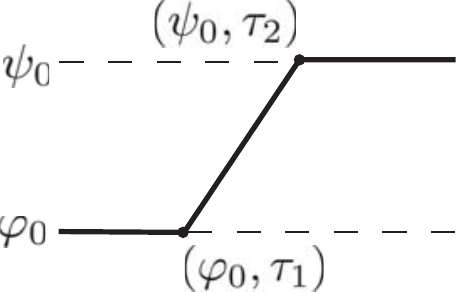}
\caption{\small{Two-slit experiment as a refraction of the electron path in $H$}}
\end{center}
\end{figure}
{\flushleft On the figure the point $(\varphi_{0},\tau_{1})$ represents the moment when the electron hits the screen with the slits. As a result of interaction with the screen, the state function of the electron in ${\widehat h}$-coordinates shifts from $(\varphi_{0}, \tau)$ to $(\psi_{0}, \tau)$. The process of passing through the slits is shown as a line segment connecting the points $(\varphi_{0},\tau_{1})$ and $(\psi_{0},\tau_{2})$. After passing the slits, the electron continues evolving as a free particle with initial state $\psi_{0}$.}

From this perspective the slits cause a refraction of the electron path in $S^{L_{2}}$. Notice the difference between Figure 6 and the standard picturing of the experiment shown in Figure 7. 
\begin{figure}[ht]
\label{fig:7}
\begin{center}
\includegraphics[width=4cm]{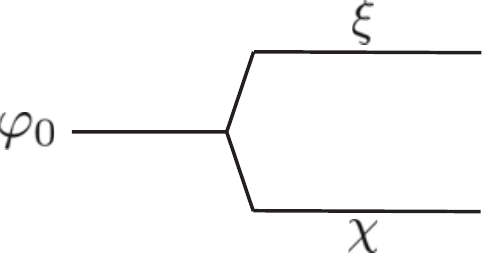}
\caption{\small{The standard picturing of the two-slit experiment}}
\end{center}
\end{figure}
The characteristic splitting of the electron path in Figure 7 is due to attaching the entire process to the classical space and is absent in Figure 6.

Assume now that a measuring device is inserted in front of one of the slits causing collapse of the electron state to, say, $\chi$. The corresponding diagram is shown in Figure 8.
\begin{figure}[ht]
\label{fig:8}
\begin{center}
\includegraphics[width=5cm]{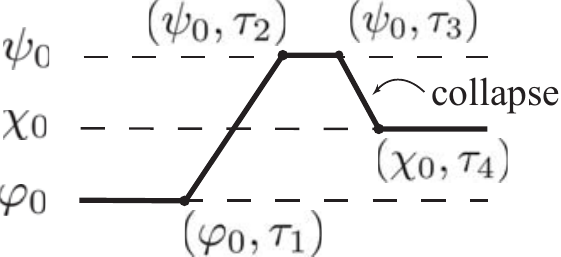}
\caption{\small{Interpretation of the two-slit experiment with collapse}}
\end{center}
\end{figure}
This simple diagram suggests that
the process of collapse in the experiment is just another refraction of the electron's path in the functional space.

To clarify this point, note that the state function of the electron is usually ``distributed'' over a range of values of its variables. At the same time, the state function is a {\em point} in the functional space $L_{2}$. In some generalized sense, the particle is a {\em point particle} in the functional space. The paradox associated with the two-slit experiment is due to the fact that we are trying to attach the process to the classical space. That is, we think of a quantum particle as being on the classical space all the time. If the process of passage through the screen is considered {\em functionally}, it can be described in terms of a simple bending of the electron's path. The same applies to the process of collapse.  

Although the mechanism of refraction of the electron path in the two-slit experiment will be treated in detail elsewhere, let us demonstrate that the ``shift'' of the path (the middle part of the diagram in Figure 6) could be indeed a geodesic in an appropriate Riemannian metric on the space of states.
For this let us consider a simpler experiment with electron in a homogeneous magnetic field. A free electron of momentum $p=\hbar k$ propagates in the direction of the $X$-axis and enters a chamber with a homogeneous magnetic field ${\bf B}=(0,B_{0},0)$. The equation of motion of the electron in the chamber is as follows:
\begin{equation}
\label{Pauli}
i\hbar \frac{d \Psi}{dt}=-\frac{\hbar^{2}}{2m}\frac{d^{2}}{dx^{2}} \Psi-\mu {\widehat \sigma_{2}}B_{0}\Psi,
\end{equation}  
where $\Psi=\Psi(s,x,t)$, $s=1,2$ is a two-components state function of the electron, $\mu$ is the electron's magnetic moment and ${\widehat \sigma}_{2}=
\left[ 
\begin{array}{cc}
0 & -i \\ 
i & 0
\end{array}
\right]$ 
is a Pauli matrix. The substitution
\begin{equation}
\Psi(s,x,t)=\psi_{t}(x)\varphi_{t}(s)
\end{equation}
produces two evolution equations. The first describes the evolution governed by the free Hamiltonian
\begin{equation}
\label{freeeH}
i\hbar \frac{d \psi_{t}}{dt}=-\frac{\hbar^{2}}{2m}\frac{d^{2}}{dx^{2}} \psi_{t}.
\end{equation}
The second equation describes the evolution in the space $C^{2}$ of spinors $\varphi$:
\begin{equation}
\label{spin_ev}
i\hbar \frac{d \varphi_{t}}{dt}=-\mu {\widehat \sigma_{2}}B_{0}\varphi_{t}.
\end{equation}
A particular solution of Eq. (\ref{Pauli}) is given by the product of the following pair of functions:
\begin{eqnarray}
\label{freeH}
\psi_{t}(x)&=&e^{i(kx-\omega t)},\\
\varphi_{t}(s)&=&\left[ 
\begin{array}{c}
\label{spinN}
cos\left(\frac{1}{2}\theta-\frac{\mu B_{0}}{\hbar}t\right) \\ 
sin\left(\frac{1}{2}\theta-\frac{\mu B_{0}}{\hbar}t\right)
\end{array}
\right],
\end{eqnarray}
where the angle $\theta$ depends on the initial spin state $\left .\varphi_{t}\right|_{t=0}\equiv \varphi_{0}$ of the electron before it enters the chamber.

Assume that $\theta=0$ so that before entering the chamber the electron is in the ``spin-up'' state, i.e., $\varphi_{0}=\left[ \begin{array}{c}
1 \\ 
0
\end{array}
\right]$.
Choose the length of the chamber in such a way that at the moment when the electron leaves the chamber it is in the spin state $\varphi_{a}=\left[ \begin{array}{c}
\frac{1}{\sqrt{2}} \\ 
\frac{1}{\sqrt{2}}
\end{array}
\right]$.
We may assume, for example, that the parameter $t$ changes between $0$ and $\frac{7\pi}{4}\frac{\hbar}{\mu B_{0}}$.
Then the process of passing through the chamber leads to a ``splitting'' of the original spin-up eigenstate of the operator $\sigma_{z}$ into a superposition of spin-up and spin-down states. In this respect the experiment is a finite dimensional version of the two-slit experiment where a localized electron wave packet gets transformed by the screen with the slits into a superposition of two wave packets. 

Let $L_{2}$ be a Hilbert space of two-component state functions and let $S^{L_{2}}$ be the sphere of unit normalized states in $L_{2}$. 
Let $M$ be the four dimensional submanifold of $S^{L_{2}}$ given by the product of manifolds
$M=I \times S^{3}$. Here $I$ is the integral curve $\psi_{t}=e^{-\frac{i}{\hbar}{\widehat h}_{0}t}\psi_{0}$ of the vector field associated with the free Hamiltonian ${\widehat h}_{0}=-\frac{\hbar^{2}}{2m}\frac{d^{2}}{dx^{2}}$ (that is, $\psi_{t}$ is a solution of Eq. (\ref{freeeH})) and $S^{3}$ is the sphere of normalized spin states. Assume for simplicity that $\psi_{0}$ is a sufficiently well localized (square-integrable) wave packet. Then the electron's path in the experiment can be described by the pair of functions $u_{t}=(\psi_{t},\varphi_{t})$, so that $u_{t}$ takes values in the submanifold $M$.  

Let us now define the Riemannian metric on the submanifold $M$ in the way consistent with Secs. 4 and 5. Namely, let ${\widehat G}=({\widehat h_{0}}{\widehat h_{0}}^{\ast})^{-1}$ be the metric on $I$ and let ${\widehat K}$ be the Killing metric on $S^{3}$. Then the Riemannian  metric on $M$ is taken to be the direct product of ${\widehat G}$ and ${\widehat K}$.   
In more detail, at each point $u=(\psi, \varphi) \in M$ the tangent space $T_{u}M$ is naturally identified with the direct sum $T_{\psi}I + T_{\varphi}S^{3}$. The metric at $u$ is then given by the block-diagonal matrix
\begin{equation}
\label{MMetric}
\left[ 
\begin{array}{cc}
{\widehat G} & 0 \\ 
0 & {\widehat K}
\end{array}
\right].
\end{equation}
As a side remark, note that the metric ${\widehat K}$ could have been written in the form analogous to ${\widehat G}=({\widehat h_{0}}{\widehat h_{0}}^{\ast})^{-1}$ (see Ref. 11).  

We claim now that the electron's path in the magnetic field is a geodesic on the manifold $M$. 
In fact, under the above assumptions the electron's path in the chamber is given by 
\begin{equation}
\label{second}
u_{t}=\left[ \begin{array}{c}
\psi_{t}\\
\varphi_{t}
\end{array}
\right]
=
\left[ \begin{array}{c}
\psi_{t}\\
cos\left(\frac{\mu B_{0}}{\hbar}t\right) \\ 
-sin\left(\frac{\mu B_{0}}{\hbar}t\right)
\end{array}
\right].
\end{equation}
We know from Sec. 4 that $\psi_{t}$ is a geodesic in the metric ${\widehat G}=({\widehat h_{0}}{\widehat h_{0}}^{\ast})^{-1}$ on $I$. Moreover, $\varphi_{t}$
is an integral curve of the left invariant vector field $\frac{i}{\hbar}\mu \sigma_{2}B_{0}$ on $S^{3}$ and is therefore a geodesic in the Killing metric (see Sec. 5). The form Eq. (\ref{MMetric}) of the metric ensures then that the curve $u_{t}=(\psi_{t},\varphi_{t})$ is a geodesic in $M$, which is what was claimed.

Let us now comment on the instantaneous nature of collapse which may find its explanation within the developed framework.
In the developed formalism
the classical space is identified with a ``spiral'' $M_{3}$ isometrically embedded into a Planck-size sphere $S^{G}$. The points on the ``spiral'' can be far apart when the distance is measured along the ``spiral''. Since the embedding $M_{3} \longrightarrow S^{G}$ is isometric, the latter distance coincides with the distance in the classical space. On the other hand, the geodesic distance between the points in the Riemannian metric on $S^{G}$ is at most of the order of radius of the sphere. 
In particular, the electron may be in a superposition $\varphi=a\chi+ b\xi$ of states of the particle localized at two distant points in space. At the same time, the functional distance between such a state $\varphi$ and the state $\chi$ (or $\xi$) may be small.
The figure below illustrates this result.
\begin{figure}[ht]
\label{fig:9}
\begin{center}
\includegraphics[width=2.6cm]{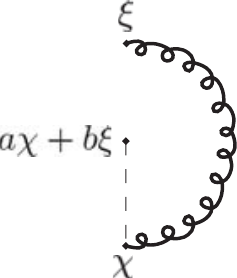}
\caption{\small{The classical space distance versus the functional distance}}
\end{center}
\end{figure}

Let us also make some comments about the dynamics of a quantum measurement. Such a dynamics is not developed in the paper. Nevertheless, there are several important observations that follow from the formalism and need to be taken into account when considering the dynamics of collapse.

First of all, the principle of functional relativity insists that, whenever valid, the Schr{\"o}dinger equation is nothing but a particular realization of a functional tensor equation
\begin{equation}
\label{inv}
\frac{d\Phi_{\tau}}{d\tau}=-i{{\bf \widehat A}}\Phi_{\tau}.
\end{equation}
Here it is assumed that ${{\bf \widehat A}}$ admits a realization as the Hamiltonian ${\widehat h}$ of the considered system.
Any other realization 
\begin{equation}
\label{ASh}
\frac{d\varphi_{\tau}}{d\tau}=-i{ \widehat A}\varphi_{\tau}
\end{equation}
of Eq. (\ref{inv}) describes a physically possible ``evolution'' in the direction specified by the operator ${\widehat A}$. 

Next, for an appropriately chosen Riemannian metric on $S^{L_{2}}$ the solution of Eq. (\ref{ASh}) through a point $\varphi_{0} \in S^{L_{2}}$ is a geodesic in the direction $-i{\widehat A}\varphi_{0}$. In particular, the evolution in an arbitrary direction of the tangent space $T_{\varphi_{0}} S^{L_{2}}$ is possible. 
Assume that the initial state $\varphi_{0}$ is an eigenstate of ${\widehat A}$ with the eigenvalue $a$. Then the equation Eq. (\ref{ASh}) is satisfied by the function 
\begin{equation}
\label{aSh}
\varphi_{\tau}=e^{-ia\tau}\varphi_{0}. 
\end{equation}
The solution Eq. (\ref{aSh}) signifies that the projection of the path $\varphi_{\tau}$ on $CP^{L_{2}}$ yields a trivial path. In other words, the eigenstates of observables are {\em zeros} of the projection of the vector field $-i{\widehat A}\varphi$ induced by the bundle projection $\pi: S^{L_{2}}\longrightarrow CP^{L_{2}}$. 

With this in hand we make the following conjecture about the nature of quantum measurement.
A classical measuring device that measures an observable ${\widehat A}$ locally curves the Riemannian metric on $S^{L_{2}}$ or $CP^{L_{2}}$. This curving results in the creation of the hole-like regions (to be called below ``holes'') on neighborhoods of the eigenstates of ${\widehat A}$ in $S^{L_{2}}$ or the corresponding points in $CP^{L_{2}}$. 
In particular, to measure position ${\widehat x}$ of a microscopic particle we may use several counters distributed in space or a photographic film. The counters or the molecules of the film play the role of the holes in $S^{G}=\left (S^{L_{2}}, G\right)$ positioned in this case along $M_{3}$, i.e., at the eigenstates of ${\widehat x}$. Similarly, to measure momentum ${\widehat p}$ of the particle, the momentum measuring devices  must be gauged in the momentum variable and play the role of holes positioned along the momentum submanifold ${\widetilde M}_{3}$ of $S^{G}$.

The evolution of a microscopic particle is a motion along a geodesics in a Riemannian metric on the sphere $S^{L_{2}}$ or on the projective space $CP^{L_{2}}$. The presence of measuring devices alters the standard Schr{\"o}dinger evolution. When the path of a particle on $S^{L_{2}}$ is close (in functional space) to a particular hole, the particle (i.e. the state!) may ``collapse'' into the hole. In particular, the state of the particle in the hole will coincide with the function that describes the position of the hole, i.e., it will be an eigenstate of the measured observable. 
The holes are zeros or ``equilibrium points'' of the vector field $-i{\widehat A}\varphi$ projected onto $CP^{L_{2}}$. The evolution of a particle in the hole is projectively trivial.  
Besides the functional distance, the collapse to a particular hole may depend on a chaotic motion of the holes (i.e. measuring molecules) along $S^{L_{2}}$. This results in a stochastic process which may account for the probabilistic character of collapse. 

Finally, let us make a brief comment about the relationship of evolutions of macroscopic and microscopic particles in the formalism. As discussed, the image of the classical space under the embedding $i: {\bf a} \longrightarrow \delta({\bf x}-{\bf a})$ is a ``spiral" through the dimensions of $S^{H}$. The standard quantum evolution of microscopic particles does not follow the ``spiral'' but rather makes a ``shortcut'' by following a geodesic of $S^{H}$. 
In particular, the microscopic particles do not normally propagate in space $M_{3}$: the path $\varphi_{\tau}({\bf x})=e^{-i{\widehat h}\tau}\varphi_{0}({\bf x})$ can hardly ever be written as a path $\delta({\bf x}-{\bf a}(\tau))$ in $M_{3}$. 
Only the particles of sufficiently large mass, or, more generally, those under a constant bombardment by the environment, are forced to stay on the classical space $M_{3}$ and evolve along the corresponding ``spiral'' in $S^{H}$. For a particle of sufficiently large mass such a motion along geodesic of $M_{3}$ can be identified with the ordinary classical motion along a straight line. Alternatively and with a good approximation the motion of sufficiently fast microscopic particles in a bubble chamber would also follow a geodesic of $M_{3}$. 

Note however, that the environment related ``bombardment'' may cause a local deformation of the metric on $S^{H}$ along the classical space $M_{3}$. In particular, $M_{3}$ may still turn out to be a totally geodesic submanifold of the sphere $S^{G}$, i.e., the sphere $S^{H}$ with an additionally deformed  metric $G$. In this case the geodesics on $M_{3}$ would also be geodesics on $S^{G}$. 
To understand how an infinitely large classical space could be embedded totally geodesically (and not only isometrically!) into an otherwise extremely small sphere $S^{G}$, one can think of the classical space in Figure 2 of Sec. 2 as a ``canyon'' on the surface of the sphere. The sphere can be small, while the ``canyon'' can be as long as one wishes, and still the curves along the bottom of the ``canyon'' could be geodesics of $S^{G}$. 
   
To become a model, the functional geometric interpretation of quantum evolution and collapse must be accompanied by the dynamical equations of motion. It was advocated here that for a single particle quantum mechanics the latter equations are simply equations of geodesics on a Hilbert Riemannian manifold. The derivation of these equations is then similar to derivation given in Secs. 4 and 5. However, the presence of measuring devices is now associated with an additional skewing of the metric. The problem is then to find the metric producing the needed geodesics. Because of that, the derivation of specific equations of collapse becomes mathematically more involved and the problem is currently open.
\bigskip

{\flushleft {\bf Acknowledgments}
\smallskip
 
I would like to thank my colleague Malcolm Forster for his faithful interest in the 
formalism, for numerous questions, comments and recommendations that helped improving many 
parts of the paper. 
I also want to express my sincere gratitude to the editor of {\it Foundations of Physics} 
for his support and understanding. }

\end{document}